\documentclass[journal]{IEEEtran}
\usepackage{cite}
\usepackage{amsmath,amssymb,amsfonts}
\usepackage{algorithmic}
\usepackage{graphicx}
\usepackage{textcomp}
\usepackage{hyperref}
\usepackage{multirow}
\usepackage{makecell}
\usepackage{rotating}
\usepackage{mathtools}
\usepackage{xspace}
\usepackage{bm} 
\usepackage{url}
\usepackage{subcaption}
\usepackage{tikz,psfrag}
\usetikzlibrary{arrows.meta}
\usetikzlibrary{arrows,positioning}
\usetikzlibrary{fit}
\usepackage[usestackEOL]{stackengine}
\tikzstyle{i} = [pin edge={to-,thin,black}]
\def\argmin#1{\underset{#1}{\text{arg min\ }}}
\def\argmax#1{\underset{#1}{\text{arg max\ }}}

\newcommand{\xmath}[1] {\ensuremath{#1}\xspace}
\newcommand{\blmath}[1] {\xmath{\bm{#1}}}
\newcommand{\A} {\xmath{\mathcal{A}}}

\newcommand{\X}{\blmath{X}}

\newcommand{\y}{\blmath{y}}

\renewcommand{\Re}{\mathbb{R}}
\newcommand{\Cp}{\mathbb{C}} 
\newcommand{\Bp}{\blmath{\phi}}
\newcommand{\BP}{\blmath{\Phi}}
\newcommand{\bv}{\blmath{v}}
\newcommand{\ba}{\blmath{\theta}}

\newcommand{\norm}[1]{\left\lVert#1\right\rVert}
\newcommand{\normr}[1]{\lVert#1\rVert} 
\newcommand{\abs}[1]{\left|#1\right|}

\providecommand{\icomplex}       {\xmath{\imath}}
\providecommand{\eexp}           {\xmath{\mathinner{\mathrm{e}}}}
\newcommand{\Expni}[1] {\xmath{\eexp^{-\icomplex #1}}}
\providecommand{\Expn}[1]       {\xmath{\eexp^{-#1}}}

\providecommand{\der}[1]        {\xmath{\mathop{\mathrm{d}#1}\nolimits}}
\newcommand{\df}{\xmath{\der f}}

\providecommand{\desa}[1]       {\begin{equation}%
    \begin{aligned}#1\end{aligned}\end{equation}} 

\begin{document}
\title{
Manifold Model for
High-Resolution fMRI \\ Joint Reconstruction and Dynamic Quantification}
\author{
Shouchang Guo, 
Jeffrey A. Fessler, \IEEEmembership{Fellow, IEEE},
and Douglas C. Noll, \IEEEmembership{Senior Member, IEEE}

\thanks{This work was supported by
National Institute of Biomedical Imaging and Bioengineering (NIBIB)
and National Institute of Neurological Disorders and Stroke (NINDS)
through grants R01 EB023618 and U01 EB026977.}
\thanks{S. Guo and J. A. Fessler are with the Department
of Electrical Engineering and Computer Science, University of Michigan, Ann Arbor, MI 48109 USA (e-mail: shoucguo@umich.edu, fessler@umich.edu).}
\thanks{D. C. Noll is with the Department of Biomedical Engineering, University of
Michigan, Ann Arbor, MI 48109 USA (e-mail: dnoll@umich.edu).}
}

\maketitle

\begin{abstract}
Oscillating Steady-State Imaging (OSSI)
is a recent fMRI acquisition method that exploits a large and oscillating signal,
and can provide high SNR fMRI.
However, the oscillatory nature of the signal leads to an increased number of acquisitions.
To improve temporal resolution and accurately model the nonlinearity of OSSI signals,
we build the MR physics for OSSI signal generation
as a regularizer for the undersampled reconstruction
rather than using subspace models that are not well suited for the data.
Our proposed physics-based manifold model
turns the disadvantages of OSSI acquisition into advantages
and enables joint reconstruction and quantification. 
OSSI manifold model (OSSIMM) outperforms subspace models
and reconstructs high-resolution fMRI images
with a factor of 12 acceleration and without spatial or temporal resolution smoothing.
Furthermore, OSSIMM can dynamically quantify important physics parameters,
including $R_2^*$ maps,
with a temporal resolution of 150 ms.
\end{abstract}


\section{Introduction}
Functional magnetic resonance imaging (fMRI) is an important tool for brain research and diagnosis. In its most common form, it detects functional activation by acquiring a time-series of MR images with blood-oxygen-level-dependent (BOLD) contrast \cite{Ogawa1990}. However, the BOLD effect has a relatively low signal-to-noise ratio (SNR) \cite{noll2001}, and the SNR further decreases with improved spatial resolution. Because the functional units (cortical columns) of the brain are on the order of 1 mm, high resolution with high SNR is critical for some fMRI experiments.
This paper focuses on Oscillating Steady-State Imaging (OSSI),
a recent fMRI acquisition approach
that provides higher SNR signals than standard gradient-echo (GRE) imaging \cite{ossi2019}. 

The SNR advantage of OSSI comes at a price of spatial and temporal resolutions. 
OSSI acquisition requires a quadratic RF phase cycling with cycle length $n_c$
(e.g., $n_c = 10$).
The corresponding OSSI signal oscillates with a periodicity of $n_c \cdot$ TR,
and the frequency-dependent oscillations result in oscillatory patterns in OSSI images.
Therefore, every image in a regular fMRI time course is acquired $n_c$ times with different phase increments in OSSI,
and combining the $n_c$ images eliminates oscillations for fMRI analysis.
Acquiring $n_c$ times more images compromises temporal resolution,
and the short TR necessary for OSSI acquisition can limit single-shot spatial resolution. 

To improve the spatial-temporal resolution,
we previously used a patch-tensor low-rank model for the sparsely undersampled reconstruction \cite{tensor}.
While low-rank regularization fits data to linear subspaces,
OSSI images are not very low-rank because of the nonlinear oscillations \cite{dictionary}.
Instead of imposing low-rankness and/or sparsity that may or may not suit the data,
this paper proposes a nonlinear dimension reduction approach for OSSI reconstruction
that uses a MR physics-based manifold as a regularizer,
inspired by parameter map reconstruction methods for MR fingerprinting \cite{Zhao2016,Asslander2018}.

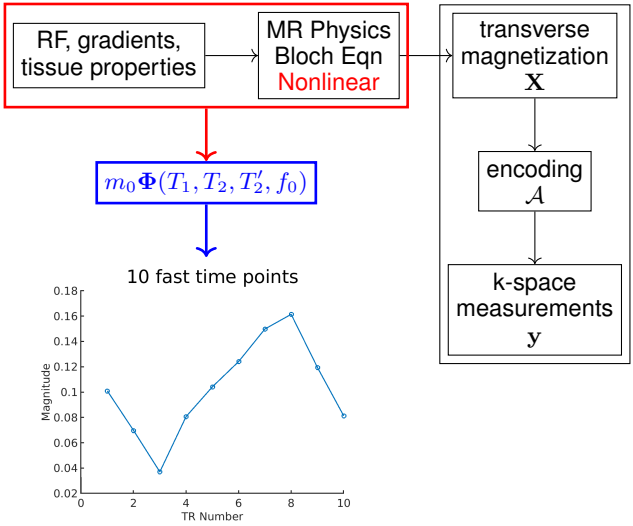
\begin{figure}
\centering
\begin{tikzpicture}[node distance=0.7cm]
\tikzset{every node}=[font=\small\sffamily]
\node [draw] (A) {\Longstack[c]{RF, gradients,\\tissue properties}};
\node [draw, right=of A] (B) {\Longstack[c]{MR Physics\\ Bloch Eqn\\ \color{red}{Nonlinear}}};
\node [draw, right=of B] (C) {\Longstack[c]{transverse \\ magnetization\\ $\mathbf{X}$}};
\node [draw, below=of C] (D) {\Longstack[c]{encoding\\ $\mathcal{A}$}};
\node [draw, below=of D] (E) {\Longstack[c]{k-space\\ measurements\\ $\mathbf{y}$}};
\draw [->] (A) edge (B) (B) edge (C) (C) edge (D) (D) edge (E);
\node[draw,fit=(C) (D) (E)] {};
\node[draw,red,line width=1pt,fit=(A) (B)] (M) {};
\node[draw,blue,line width=1pt,below=of M] (N) {$m_0 \mathbf{\Phi} (T_1, T_2, T_2', f_0)$};
\path[->,red,line width=1pt] (M) edge (N);
\node [draw, below=of N, white] (F) {\includegraphics[height=0.14\textheight]{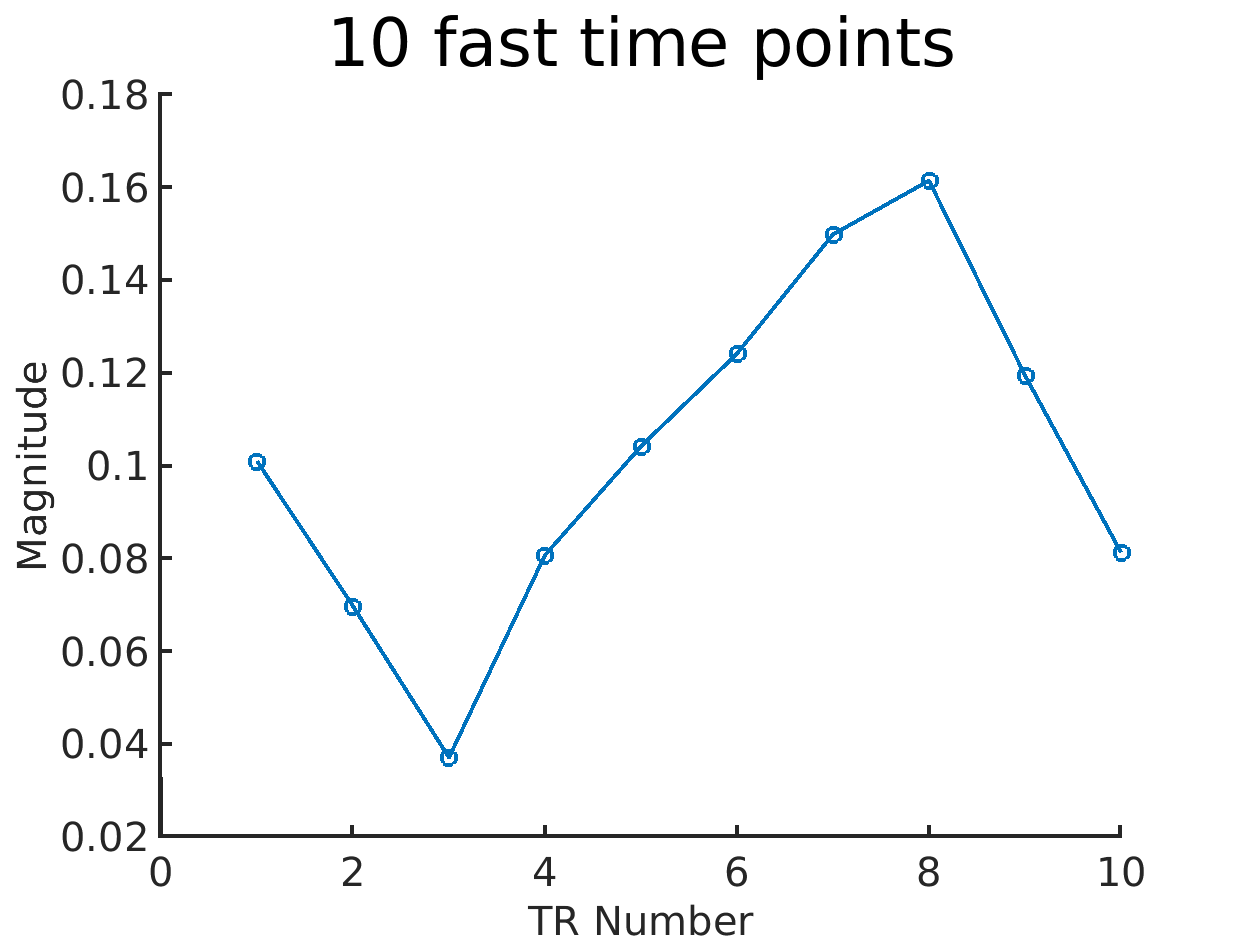}};
\path[->,blue,line width=1pt] (N) edge (F);
\end{tikzpicture}
\caption{
The proposed manifold model uses the MR physics for signal generation
as a regularizer for the undersampled reconstruction.
}\label{mf0}
\end{figure}

As outlined in Fig.~\ref{mf0},
the manifold model focuses on MR physics for OSSI signal generation.
It represents $n_c$ OSSI signal values per voxel
by just 3 physical parameters, via Bloch equations.
The nonlinear nature of the Bloch equations
enables nonlinear representations of the data and nonlinear dimension reduction.
We further introduce a near-manifold regularizer
that 
encourages the reconstructed signal values
to lie near the manifold.
Compared to quantitative imaging works that enforce the reconstructed images
to be exactly equal to the physics-based representations
\cite{Zhao2016,Asslander2018,Dong2019,Tamir2020},
the proposed near-manifold regularizer encourages the images to be near the manifold
while also allowing for potential model mismatch.

Standard $T_2^*$-weighted magnitude images only assess relative signal changes due to BOLD effects
and are not quantitative in terms of the blood oxygenation level, $T_2^*$ or $T_2’$ \cite{Speck1998,Wennerberg2001,Olafsson2008}.
Quantifying $T_2^*$ is important because of its sensitivity to iron concentration for disease monitoring \cite{Wang2019}.
By constructing a $T_2’$ manifold based on BOLD-induced intravoxel dephasing,
our work demonstrates the utility of the OSSI manifold model
for dynamic quantification of $T_2^*$/$R_2^*$.

This paper shows that the proposed $T_2’$ manifold and near-manifold regularizer
can jointly optimize OSSI images and quantitative maps.
The manifold model enables high-resolution OSSI fMRI with 12-fold acquisition acceleration,
outperforms low-rank regularization with more functional activation,
and provides quantitative and dynamic assessment of tissue $R_2^*$ maps
and off-resonance $f_0$,
with a temporal resolution of 150~ms.

\section{OSSI Manifold Model (OSSIMM)}
OSSI signal oscillates with a periodicity of $n_c$TR, and the OSSI fMRI time course contains $n_c$ images for every image in a regular fMRI time series. We refer to the fast acquisition dimension of size $n_c$ as ``fast time’’ and the regular fMRI time dimension as ``slow time’’ as presented in supplemental Fig.~\ref{msf0}.
OSSI fast time signals can have different shapes
and change nonlinearly
with respect to MR physics parameters,
as illustrated in Fig. \ref{mf1}.
To accurately model the nonlinear oscillations,
we propose a MR-physics based manifold model for the undersampled reconstruction.

\subsection{Physics-Based Manifold}

In OSSI, the steady-state transverse magnetization of one isocromat at observation time $t$ is
\[
m_0 \, \Bp (t; T_1, T_2, f_0), 
\]
where $m_0 \in \Cp$ is the equilibrium magnetization,
$\Bp(\cdot) \in \Cp^{n_c}$ represents MR physics calculated by Bloch equations,
$T_1$ and $T_2$ are tissue relaxation times,
and $f_0$ denotes central off-resonance frequency from $B_0$ field inhomogeneity.

$T_2’$-weighted OSSI signal in a voxel with an intra-voxel spreading of off-resonance frequencies $f$ can be modeled as:
\desa{
\begin{split}
m_0  & \BP (t; T_1, T_2, T_2', f_0) = \\
& \int m_0 \Bp (t; T_1, T_2, f_0 + f) \, p(f; T_2') \df
.
\end{split}
}
The $T_2’$ exponential decay corresponds to a Cauchy distribution
for $f$ with a probability density function (PDF)
$p(f) = \gamma / \pi (\gamma^2 + f^2)$, and scale parameter $\gamma = 1 / (2 \pi T_2')$.

The isocromat signal at time $t > 0$ presents increased $T_2$ decay and increased off-resonance dephasing due to field inhomogeneity and BOLD-related field changes,
\desa{
\begin{split}
m_0 & \Bp (t; T_1, T_2, f_0) = \\
& m_0 \Bp (t = 0; T_1, T_2, f_0) 
\Expn{t / T_2}
\Expni{2\pi f_0 t}
,
\end{split}
}
where $t = 0$ denotes the time right after the excitation.


As OSSI TR is relatively short (e.g., TR = 15 ms),
we neglect the intravoxel dephasing during the readout
and approximate the signal at $0 \leq t \leq \textrm{TR}$
with the signal at the echo time TE.
The $T_2’$-weighted signal becomes
\desa{
\begin{split}
m_0 \BP (T_1, T_2, T_2', f_0)
& \approx \int m_0 \Bp (\textrm{TE}; T_1, T_2, f_0 + f) 
\\&
\Expn{\textrm{TE}/ T_2}
\Expni{2\pi (f_0 + f)\textrm{TE}}
p(f; T_2') \df
\label{T2s}
.
\end{split}
}

Accordingly, $T_2’$-weighted OSSI fast time signals lie on the physics-based manifold:
\desa{ 
\{m_0 \BP (T_1, T_2, T_2', f_0) \in \Cp^{n_c}:
m_0 \in \Cp, \, T_1, T_2, T_2', f_0 \in \Re
\},
}
The manifold maps a limited number of physics parameters to the $n_c$-dimensional oscillating signals via MR physics. 

\subsection{Near-Manifold Regularization}

\begin{figure}
    \includegraphics[width=0.48\textwidth]{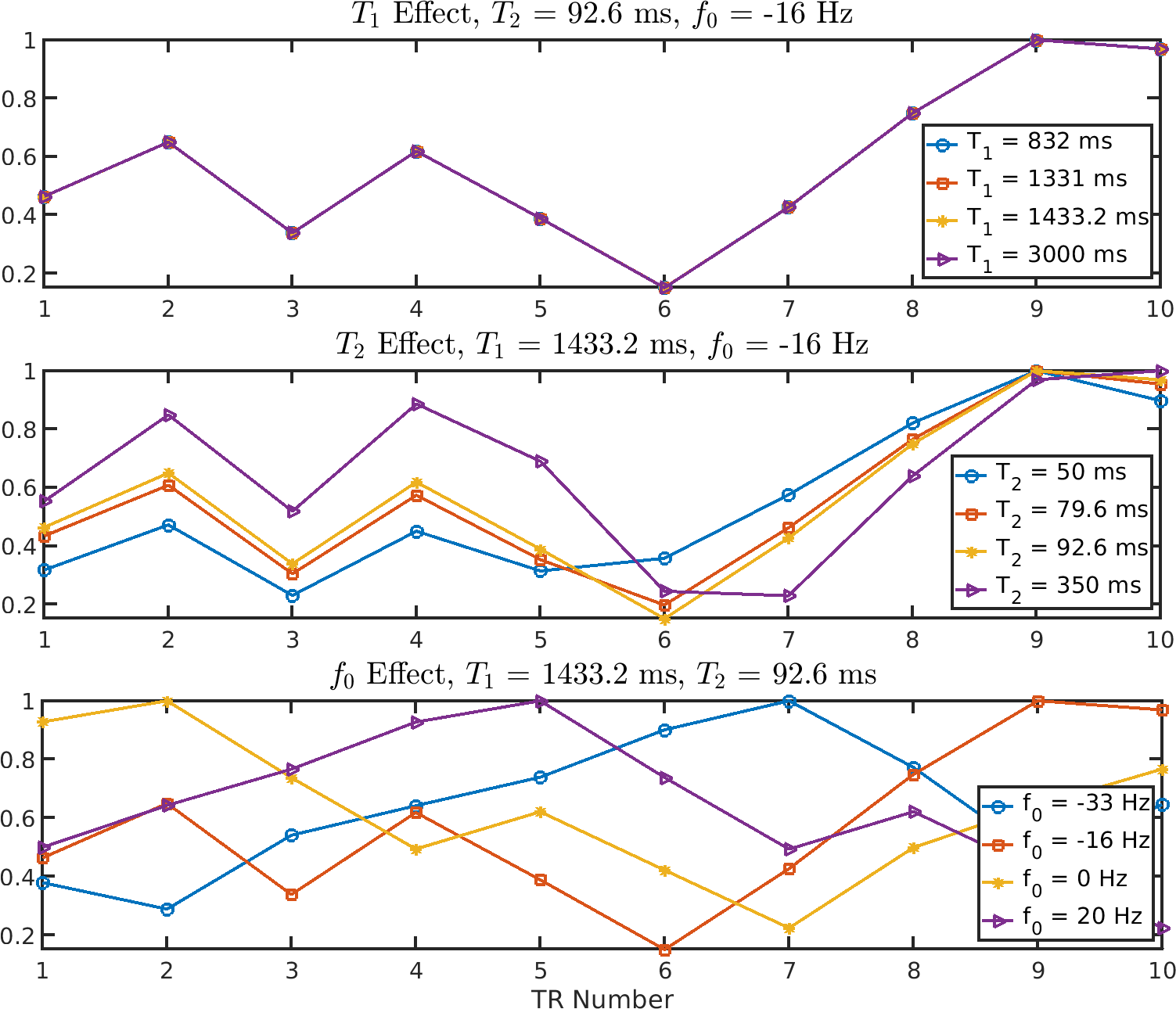}
   \caption{
   Normalized OSSI fast time signal magnitude for one isocromat with nonlinear oscillations determined by physics parameters $T_2$ and $f_0$. The change of $T_1$ only scales OSSI signal values.
   }\label{mf1}
\end{figure}

The physics-based manifold models the generation of MR signals,
enables nonlinear dimension reduction,
and can be an accurate prior for the undersampled reconstruction. 
Because the physics parameters are location dependent,
and because OSSI signal values change drastically with varying parameters as shown in Fig.~\ref{mf1},
we model the fast time signals in a voxel-by-voxel manner.
Furthermore, to account for potential mismatches due to model simplifications
and nonidealities in experiments (e.g., flip angle inhomogeneity),
we propose a near-manifold regularizer that encourages the signal values in each voxel
to be close to the manifold estimates but not necessarily exactly the same. 

The proposed $T_2’$ manifold-based image reconstruction problem
uses the following optimization formulation:
\desa{
\begin{array}{ll}
\hat{\X} =
\argmin{\X}
\frac{1}{2}\lVert\mathcal{A}(\X)-\y \rVert_2^2
+ \beta \sum_{n=1}^N \mathcal{R} \left(\X[n,:]\right)
, \quad
\\[1.5em]
\mathcal{R}(\bv) = \underset{m_0,T_2', f_0}{\textrm{min}}\ \lVert \bv-m_0 \BP (T_2', f_0; T_1, T_2)\rVert_2^2,
\end{array}
\label{costmf}
}
where
$\X \in \Cp^{N \times n_c}$
denotes $n_c$ fast time images to be reconstructed. The vectorized spatial dimension $N$ is $N_{xy}$ for 2D OSSI fMRI. 
$\mathcal{A}(\mathbf{\cdot})$ is a linear operator consisting of coil sensitivities and the non-uniform Fourier transform including undersampling,
$\y$ represents sparsely sampled k-space measurements.
$\beta$ is the regularization parameter.
$\bv \in \Cp^{n_c}$ is a vector of fast time signal values for each voxel in $\X$, 
$m_0\BP(T_2', f_0; T_1, T_2) \in \Cp^{n_c}$ denotes the manifold estimates.
The regularizer minimizes the Euclidean distance between $\bv$ and $m_0\BP(T_2', f_0; T_1, T_2)$.
$T_1$ and $T_2$ are not directly estimated by the model.
$T_1$ has a signal scaling effect
that can be absorbed in $m_0$,
as illustrated in Fig.~\ref{mf1}.
Section~\ref{sec:simu} describes
the choices of baseline $T_2$ values for $T_2^*$ estimation.

The voxel-wise parametric regularizer $\mathcal{R}(\bv)$
not only performs regularization for the ill-posed reconstruction problem,
but also involves parameter estimation and can provide quantitative maps for $T_2'$ and $f_0$. 

\subsection{Optimization Algorithm} 

To solve
\eqref{costmf},
we alternate between a regularization update and a data fidelity update for the reconstruction. 
The minimization of the voxel-wise parametric regularizer is a nonlinear least-square problem
that we solve using the variable projection (VARPRO) method \cite{Golub1973,Golub2003}. 
Let $\ba = [T_2', f_0]$ denote the two nonlinear tissue parameters;
the calculation of $\ba$ using VARPRO simplifies to
\desa{
\hat{\ba} 
= \argmax{\ba}
\frac{\abs{{\BP(\ba)}'\bv}^2}{\norm{\BP(\ba)}_2^2},
\label{up1}
}
where
$\bv = \X[:,n] \in \Cp^{n_c}$.
Instead of solving \eqref{up1} for the explicit and sophisticated $\BP(\ba)$, 
we construct a dictionary consisting of discrete $\BP(\ba)$ realizations
with varying $\ba$ parameters using Bloch simulations,
and then perform grid search
to find $\hat{\ba}$
for which $\BP(\hat{\ba})$ best matches $\bv$. 

Updating $m_0$ is a least-squares problem with closed-from solution:
\desa{
\hat{m}_0 = \frac{{\BP(\hat{\ba})}'\bv}{\normr{\BP(\hat{\ba})}_2^2}
\label{up2}
.}
We parallelize the regularization update across different voxels.

The update step for \X
involves a quadratic least-squares problem
that we solve using the conjugate gradient method
as implemented in the Michigan Image Reconstruction Toolbox \cite{fesslermirt}.
This data fidelity update is easily parallelized over different fast time images or different fast time images sets
to speed up the fMRI time series reconstruction. 

\subsection{Comparison Method}

We compare the manifold approach to a low-rank reconstruction approach
that models the fast time signals using linear subspaces.
The cost function for this low-rank comparison method is
\desa{
\begin{array}{ll}
\hat{\X} =
\argmin{\X}
\frac{1}{2}\lVert\mathcal{A}(\X)-\y \rVert_2^2
+ \alpha \lVert \X \rVert_*
\end{array}
\label{costlr}
}
where
$\X \in \Cp^{N \times n_c}$ represents every $n_c$ fast time images,
and $\alpha$ is the regularization parameter.
We solve the optimization problem \eqref{costlr}
using the proximal optimized gradient method (POGM)
with adaptive restart
\cite{kim2018adaptive,Taylor2017,lin:19:edp}.

\section{Simulation Investigations}
\label{sec:simu}
We generated OSSI signals via Bloch simulation
using pulse-sequence parameters that matched the actual data acquisition.
We used TR = 15 ms, TE = 2.7 ms (spiral-out trajectory), RF excitation pulse length = 1.6 ms,
quadratic RF phase cycling with $\Phi(n) = {\pi n^2}/{n_c}$ for $n$th TR, $n_c = 10$,
and flip angle = $10^\circ$ \cite{ossi2019}. 

\begin{figure*}
     \centering
     \subcaptionbox{\label{mf2a}}{\includegraphics[width=.48\textwidth]{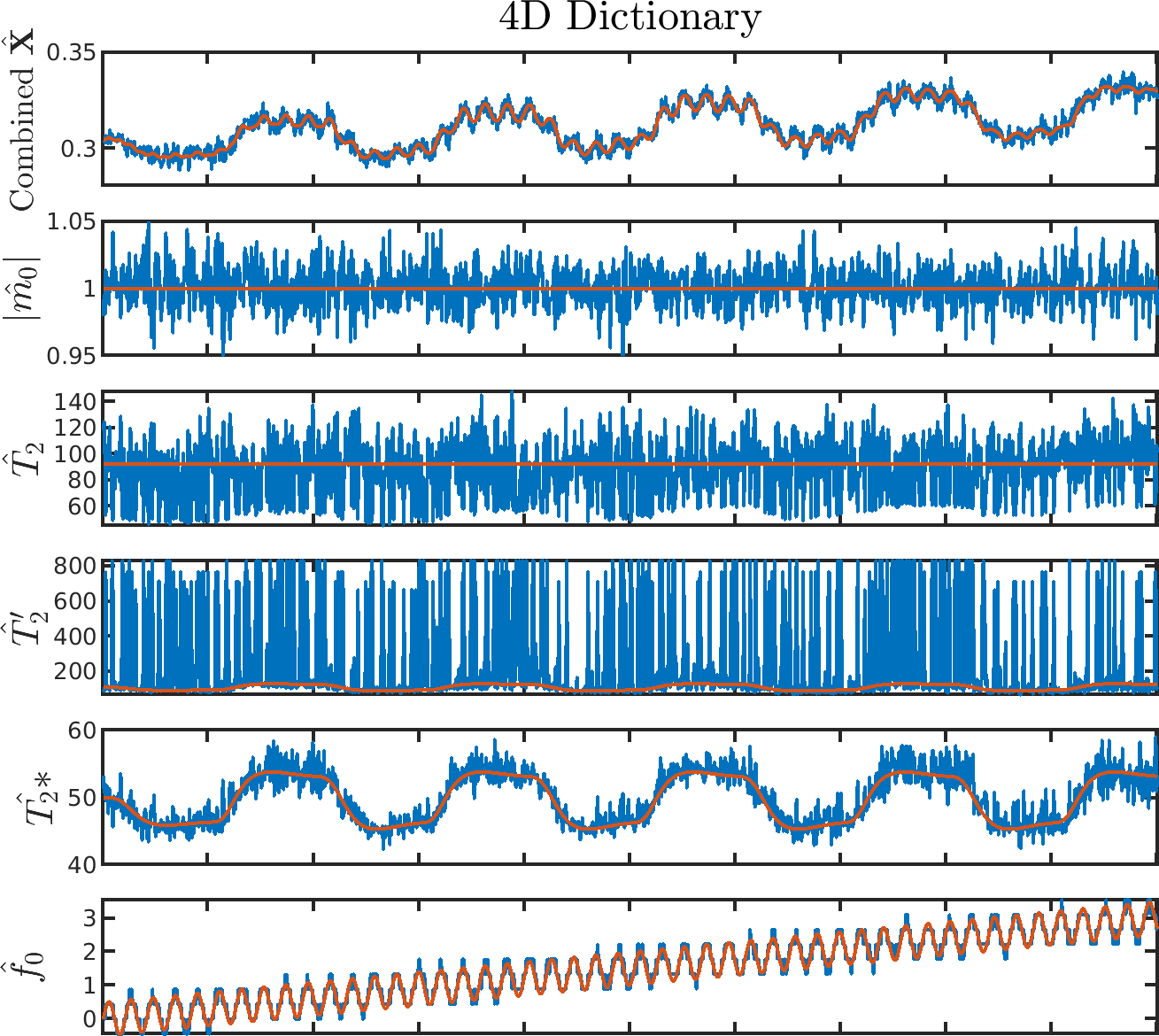}}
     \subcaptionbox{\label{mf2b}}{\includegraphics[width=.48\textwidth]{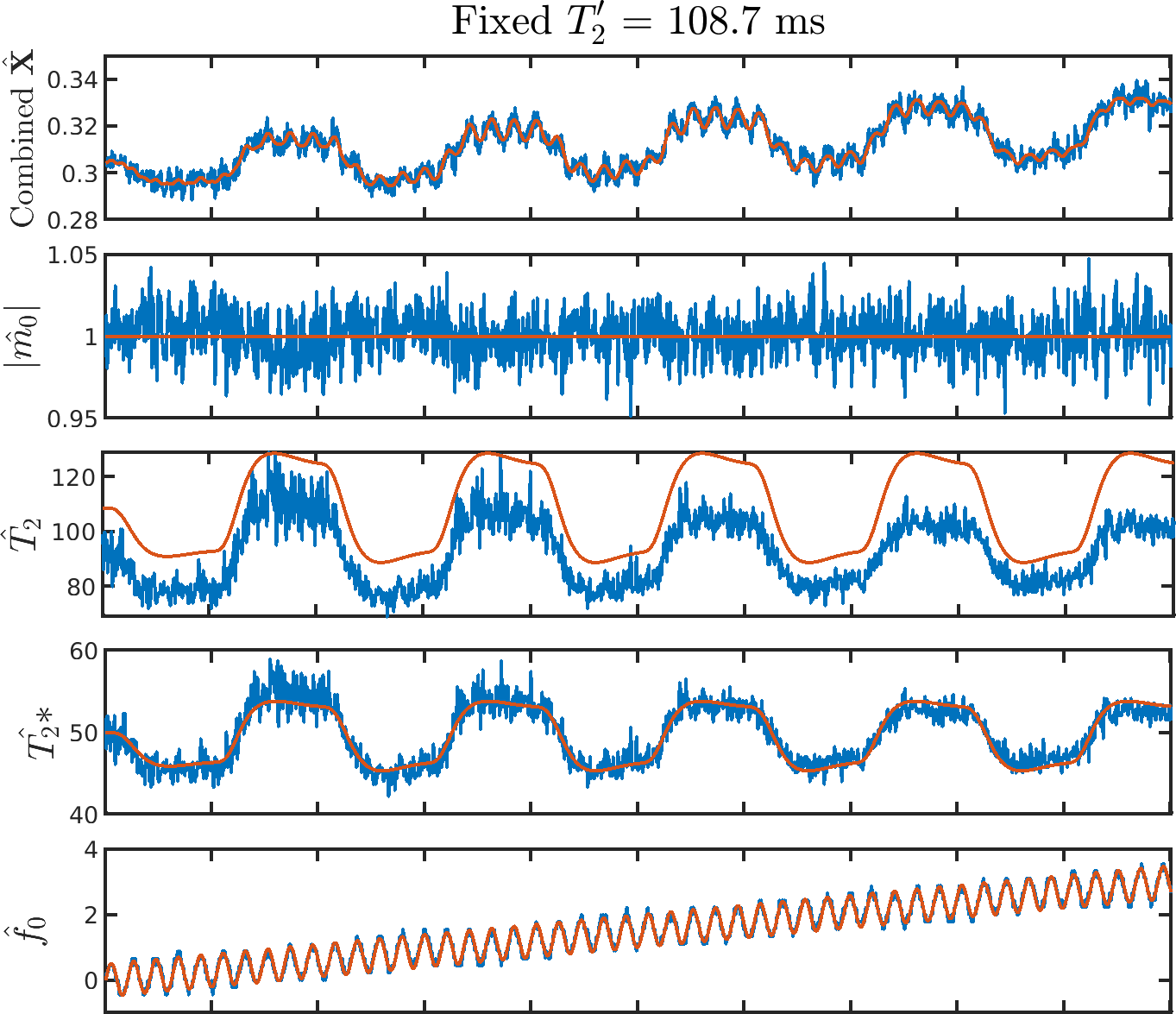}}
     \subcaptionbox{\label{mf2c}}{\includegraphics[width=.48\textwidth]{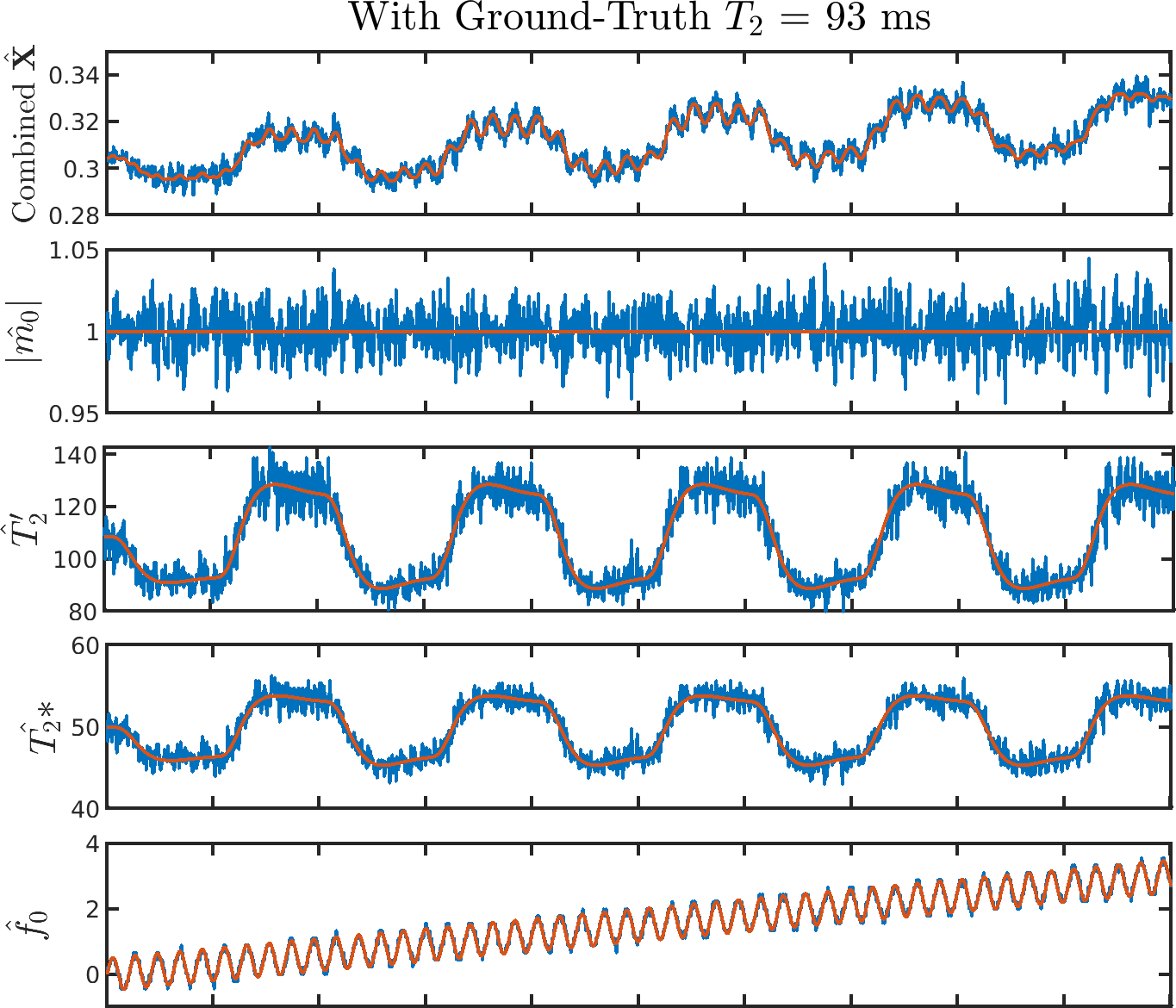}}
     \subcaptionbox{\label{mf2d}}{\includegraphics[width=.48\textwidth]{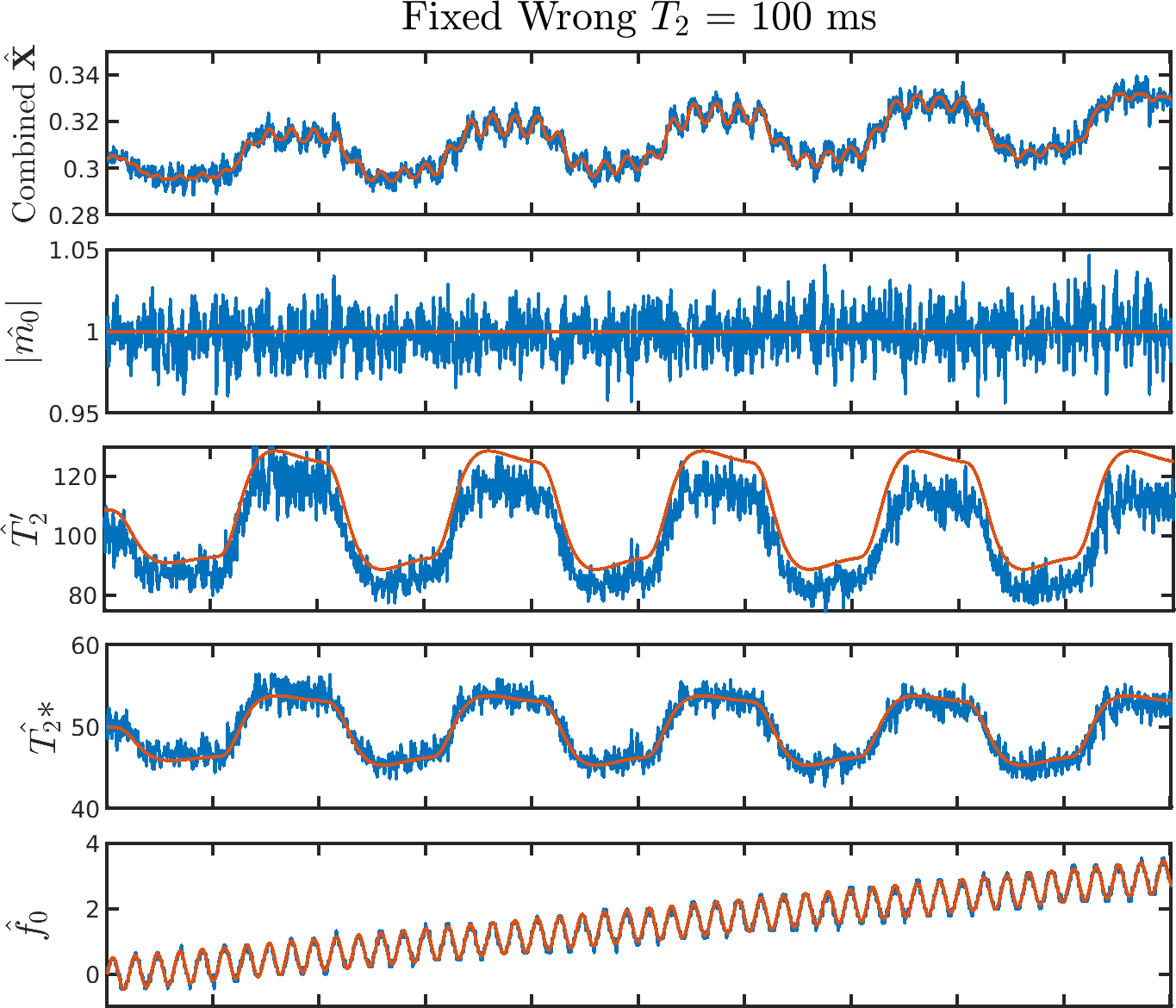}}
     \caption{
     Quantification results for a simulated OSSI fMRI voxel using the manifold model with 4 different choices of the manifold. Because $T_2$ and $T_2'$ effects to OSSI signals are correlated (Fig.~\ref{mf2a}), and a $T_2$ manifold is not good enough for capturing BOLD-induced $T_2'$ changes (Fig.~\ref{mf2b}), we use a $T_2'$ manifold for quantification. We can estimate $T_2^*$ and $T_2'$ with known $T_2$ values (Fig.~\ref{mf2c}), or use a biased guess of $T_2$ for quantifying $T_2^*$ (Fig.~\ref{mf2d}).
     }
     \label{mf2}
\end{figure*}

\subsection{OSSI Signals}

The OSSI signal $\in \mathbb{C}^{n_c}$ for one isocromat is determined by physics parameters $T_1$, $T_2$, and $f_0$. 
Fig. \ref{mf1} presents example OSSI isocromat signals (normalized by the maximum magnitude)
with varying physics parameters selected based on gray matter relaxation parameters:
$T_1$ = 1400 ms, $T_2$ = 92.6 ms \cite{Parameters}. 
As an approximation of \eqref{T2s},
we simulated $T_2’$-weighted OSSI signal in a voxel with Riemann sum of numerous OSSI isocromat signals
at different off-resonance frequencies.
Specifically, we calculated a weighted sum of OSSI signals
from 4000 isocromats at off-resonance frequency $f_0+f$,
where $f$ uniformly ranged from -200 Hz to 200 Hz,
and the weighting function was the PDF of the Cauchy distribution.

We further simulated a fMRI time course for one voxel with time-varying $T_2’$ values. 
The $T_2’$ waveform is the convolution of the canonical hemodynamic response function (HRF) \cite{spm}
and the fMRI task waveform. 
Because fMRI percent signal change
$\Delta \% \approx \Delta R_2’ \cdot \textrm{TE}_\textrm{eff}$ \cite{Jin2006}
and OSSI $\textrm{TE}_\textrm{eff}$ = 17.5 ms \cite{ossi2019},
we set $\Delta T_2’$ = 15.4 ms to produce a typical percent signal change of 2\%.
The fMRI time course is also affected by scanner drift and respiration induced $f_0$ changes.
We simulated $f_0$ with a linearly increasing scanner drift of about 1 Hz per minute
and a sinusoidal waveform (magnitude of 0.5 Hz and period of 4.2 s) to model the respiratory changes.
We also added complex Gaussian random noise for a typical temporal SNR (tSNR) value of 38 dB.

\subsection{Dictionary Selection}

We represented OSSI manifold using a signal dictionary,
and each dictionary atom is a point on the manifold.
Because $T_2$, $T_2’$, and $f_0$ affect OSSI signals in different ways
while $T_1$ has a scaling effect,
we constructed a 4D dictionary by varying $T_2$, $T_2’$, $f_0$,
for $T_1$ = 1400 ms.
The $T_2$ grids were in the 40 to 150 ms range with a 1 ms spacing.
The $T_2’$ grids were calculated by uniformly changing $R_2^*$ from 12 to 38 Hz \cite{Peran2007}
with a step size of 0.1 Hz and a fixed $T_2$ of 92.6 ms.
We set central off-resonance frequency $f_0$ to [-33.3,33.3] Hz
with a 0.22 Hz spacing as OSSI signals are periodic
with off-resonance frequency period = 1/TR = 66.7 Hz \cite{ossi2019}.

We reconstructed the functional signal and physics parameters from the simulated noisy fMRI time courses
using the near-manifold regularizer in \eqref{costmf} and the 4D dictionary.
The reconstructions were performed by
(a) simultaneously estimating $T_2$ and $T_2’$ using the 4D dictionary,
(b) assuming $T_2’$ is fixed and estimating $T_2$
using the 3D subset of the 4D dictionary based on the assumed $T_2’$ value,
(c) estimating $T_2’$ with the actual $T_2$ value and the corresponding 3D dictionary,
(d) assuming $T_2$ is fixed and estimating $T_2’$ with a biased $T_2$ value and the corresponding 3D dictionary.

As shown in Fig.~\ref{mf2},
because of the strong coupling between $T_2$ and $T_2’$ values,
it is infeasible to simultaneously estimate $T_2$ and $T_2’$ (see Fig.~\ref{mf2a}).
Using a biased $T_2’$ value for $T_2$ estimation (Fig.~\ref{mf2b})
or a biased $T_2$ value for dynamic $T_2’$ estimation (Fig.~\ref{mf2d})
results in noticeable bias,
whereas Fig.~\ref{mf2c} presents accurate $\hat{T}_2'$ when the ground truth $T_2$ is provided. 
However, all the different estimation approaches lead to relatively good $T_2^*$ estimates.
Because $m_0$ and $T_2^*$ estimates are more accurate in Figs.~\ref{mf2c} and~\ref{mf2d},
we propose to use assumed $T_2$ values
or to measure accurate baseline $T_2$ maps
to use for dynamic $T_2^*$ quantification.
The latter approach also provides $T_2’$ estimates.
Notably,
the quality of the combined functional signals is insensitive to the choice of manifold for reconstruction.

\section{Experiments}
We collected resolution phantom data and human fMRI data to evaluate the potential of the manifold model for joint reconstruction and quantification.
All the data were acquired with a 3T GE MR750 scanner (GE Healthcare, Waukesha, WI) and a 32-channel head coil (Nova Medical, Wilmington, MA). 

\subsection{Data Acquisition}

OSSI acquisition parameters were the same as in Simulation Investigations
with 10~s discarded data points to ensure the steady state.
We selected a 2D oblique slice passing through the visual cortex
with FOV = $220 \times 220 \times 2.5 \, \text{mm}^3$,
matrix size = $168 \times 168 \times 1$,
and spatial resolution = $1.3\times 1.3 \times 2.5 \, \text{mm}^3$. 
For OSSI, both ``mostly sampled’’ data (for retrospective undersampling) and prospectively undersampled data were acquired.
The sampling trajectories were undersampled VD spirals with golden-angle based rotations between time frames
as in \cite{tensor}. 
The ``mostly sampled’’ data used number of interleaves $n_i = 9$ VD spirals with approximately a 1.5 undersampling factor,
and temporal resolution = 1.35 s = $\text{TR} \cdot n_c \cdot n_i$.
The retrospective undersampling used the first interleave out of 9 for each time frame of the ``mostly sampled’’ data.
The prospective undersampling used $n_i = 1$ with temporal resolution = 150 ms = $\text{TR} \cdot n_c$
Both retrospective and prospective undersampling provided $12\times$ acceleration.

For quantification evaluation, we acquired multi-echo GRE images to get standard estimations of $f_0$ and $R_2^*$ values. 
GRE images were collected with a spin-warp sequence with TR = 100 ms, Ernst flip angle = 16$^\circ$, and different TEs = 5.9, 13, 26, and 40 ms.
$R_2^*$ maps were estimated based on the exponential decay of $T_2^*$. 
The field map $f_0$ was estimated using fully sampled GRE images at TE = 30 and 32 ms \cite{noll1995spiral}. 
For the phantom data, we additionally acquired spin-echo images with a spin-warp sequence at TR = 400 ms and different TEs = 20, 40, 60, and 80 ms to get $\hat{T}_2$ maps.  

For coil sensitivity map calculation, we collected spin-warp images and generated ESPIRiT sensitivity \cite{Uecker2014,Bart} after compressing the 32-channel coil images to 16 virtual coils using PCA \cite{Huang2008AChannels}. The coil images were 2-norm combined for brain region extraction using the Brain Extraction Tool \cite{Smith2002}.

For human data, the functional task was a left vs. right reversing-checkerboard visual stimulus with 10 s rest followed by 5 cycles of left or right stimulus (20 s L/20 s R $\times$ 5 cycles). The 10 s resting-state data ensured the oscillating steady state and were discarded. The number of time frames (both fast time $n_c$ and slow time) was 1490 for ``mostly sampled’’ data and was 13340 for prospectively undersampled data. 

\subsection{Performance Evaluation}

Every non-overlapping set of $n_c = 10$ fast time images were reconstructed and 2-norm combined for fMRI analysis.
To avoid modeling error from the HRF of the initial rest period, the data for the first 40 s task block were discarded.
The data were detrended using the first 4 discrete cosine transform basis functions to reduce effects of scanner drift.

We evaluated the functional performance of OSSIMM and comparison approaches using activation maps and tSNR maps. 
The backgrounds of activation maps were the mean of time-series of images.
The activated regions of activation maps were determined by correlation coefficients above a 0.45 threshold.
The correlation coefficients were generated by correlating the reference waveform (task and HRF related)
with the fMRI time course for each voxel.
For each voxel, dividing the mean of the time course by the standard deviation of the time course residual
(mean and task removed) provided the tSNR map.
We further calculated numbers of activated voxels at the bottom third of the brain
(where the visual cortex is located) and the average tSNR values within the brain (after skull stripping).

For quantification, parameter estimations at regions with little or no signal are masked out.
Specifically, we generated a mask with the first-echo GRE image (TE = 5.9 ms and after skull stripping)
for signals larger than 10\% of the signal magnitude and GRE $\hat{R}_2^*\, <$ 50 Hz.
Regions with GRE $\hat{R}_2^*\, >$ 50 Hz are concentrated at the edge of the brain as shown in Fig.~\ref{msf1}. 
The quantitative accuracy of OSSI $\hat{R}_2^*$ was evaluated by RMSE with multi-echo GRE $\hat{R}_2^*$ as the standard.
Because OSSI $\hat{f}_0$ estimates are in the range of [-33.3, 33.3] Hz,
we mapped the GRE $\hat{f}_0$ to the same range for comparison.

\begin{figure}
\centering
     \includegraphics[angle=-90,origin=c,width=0.49\textwidth]{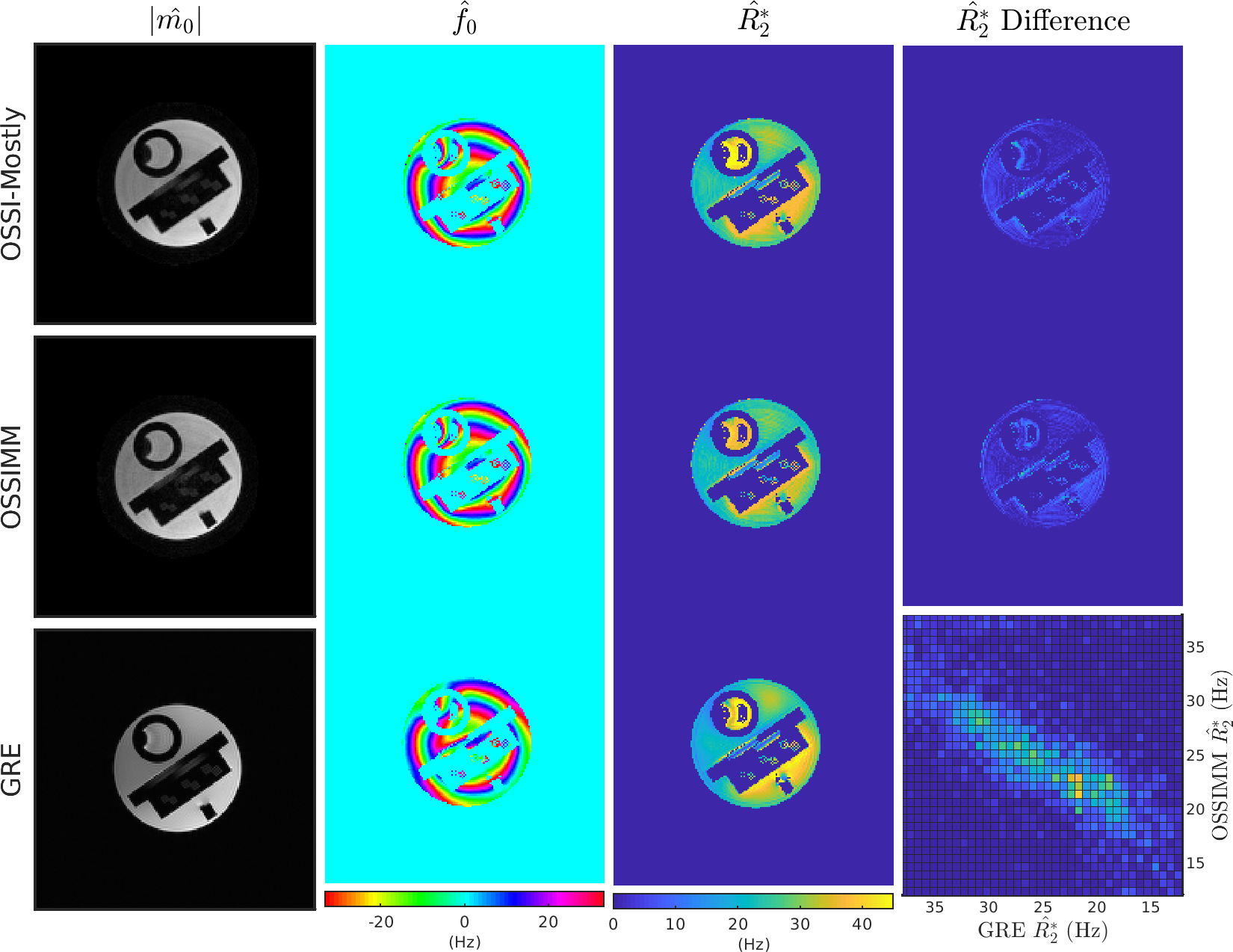}
   \caption{
   Phantom quantification of $m_0$, $f_0$, and $R_2^*$ from mostly sampled OSSI data, retrospectively undersampled OSSI data (reconstructed and quantified using OSSIMM), and multi-echo GRE. The $\hat{m}_0$ estimates are on arbitrary scales. The GRE $\hat{R}_2^*$ map is used as the standard for difference map calculation. The $\hat{R}_2^*$ maps and $\hat{R}_2^*$ difference maps use the same color scale. The 2D histogram (bottom right) compares OSSIMM and GRE $\hat{R}_2^*$ within the 12-38 Hz range. OSSI $\hat{R}_2^*$ and GRE $\hat{R}_2^*$ demonstrates similar contrasts.
   }\label{mf3}
\end{figure}

\section{Reconstruction, Quantification, and Results}
The proposed OSSIMM method jointly reconstructed high-resolution images and quantitative maps
using the near-manifold regularizer.
For both phantom and human experiments,
we used the $T_2'$ manifold with a fixed $T_2 = 100$~ms unless otherwise specified.
After reconstructing fast time images with mostly sampled data (OSSI-Mostly)
, or other models such as low-rank (OSSI-LR) and regularized cgSENSE (OSSI-cgSENSE),
we further estimated their parameter maps via dictionary fitting using the same manifold as in OSSIMM. 

\subsection{Implementation Details}

We selected the regularization parameters
based on the spectral norm $\sigma(\A)$ calculated with power iteration.
We set the regularization parameter $\beta$ in \eqref{costmf} to be a fraction of $\sigma(\A)$
such that the condition number of the cost function was about 10 to 20
and the performance of functional maps is maximized.
We selected $\alpha$ in \eqref{costlr} to enforce that matrix rank of $\hat{\X}$ was $\approx$ 4
for most fast-time image sets
and to maximize the functional performance. 

In OSSIMM, we used 4 iterations of alternating minimization,
and 2 iterations of conjugate gradient for the data fidelity update.
We used 15 iterations of POGM for the LR reconstruction
and 19 iterations of conjugate gradient for cgSENSE reconstruction of undersampled and mostly sampled data.
We generated data-shared images as initialization for the undersampled reconstructions
by using the sampling incoherence between fast and slow time \cite{tensor}
and combining k-space data of every 10 slow time points.

\subsection{Results}

\begin{table}
\renewcommand{\arraystretch}{1.6}
\caption{Phantom quantification comparison of OSSI $\hat{R}_2^*$ to GRE with or without a known $\hat{T}_2$ map}
\label{mt1}
\centering
\begin{tabular}{ccccc}
\hline\hline
\multirow{2}{*}{} & \multicolumn{2}{c}{Fixed $T_2$ = 100 ms} & \multicolumn{2}{c}{Known $\hat{T}_2$ map} \\ \cline{2-5} 
& \begin{tabular}[c]{@{}c@{}}$\hat{R}_2^*$ RMSE \\[-0.6em] (Hz)\end{tabular}
& \begin{tabular}[c]{@{}c@{}}Additional \\[-0.6em] Mask\end{tabular} 
& \begin{tabular}[c]{@{}c@{}}$\hat{R}_2^*$ RMSE \\[-0.6em] (Hz)\end{tabular}
& \begin{tabular}[c]{@{}c@{}}Additional \\[-0.6em] Mask\end{tabular}
\\ \hline 
OSSI-Mostly & 4.9 & 4.3 & 5.0 & 4.6           
\\ \hline
OSSIMM & 5.5 & 4.6 & 5.3 & 4.5             
\\ \hline\hline
\end{tabular}
\end{table}

For the phantom study,
Fig.~\ref{mf3} and Fig.~\ref{msf2} present OSSI quantification results
with a fixed $T_2$ of 100 ms and a known $\hat{T}_2$ map, respectively.
OSSIMM quantifies parameters from retrospectively undersampled data,
and results in similar maps as mostly sampled reconstruction and multi-echo GRE.
The 2D histogram demonstrates a close to a linear relationship between OSSI and GRE $\hat{R}_2^*$ values.
As summarized in Table~\ref{mt1},
OSSIMM with a known $\hat{T}_2$ map produces similar results as OSSIMM with a fixed $T_2$ value.
Demonstrated by RMSE values with additional masking in Table \ref{mt1}, OSSI $\hat{R}_2^*$ RMSE improves by 0.5-1 Hz when a GRE $12 < \hat{R}_2^* < 38$ mask (within OSSIMM $R_2^*$ dictionary range) is applied.

\begin{figure*}
  \begin{minipage}[t]{0.7\textwidth}
  \vspace{0pt}
    \includegraphics[width=\textwidth]{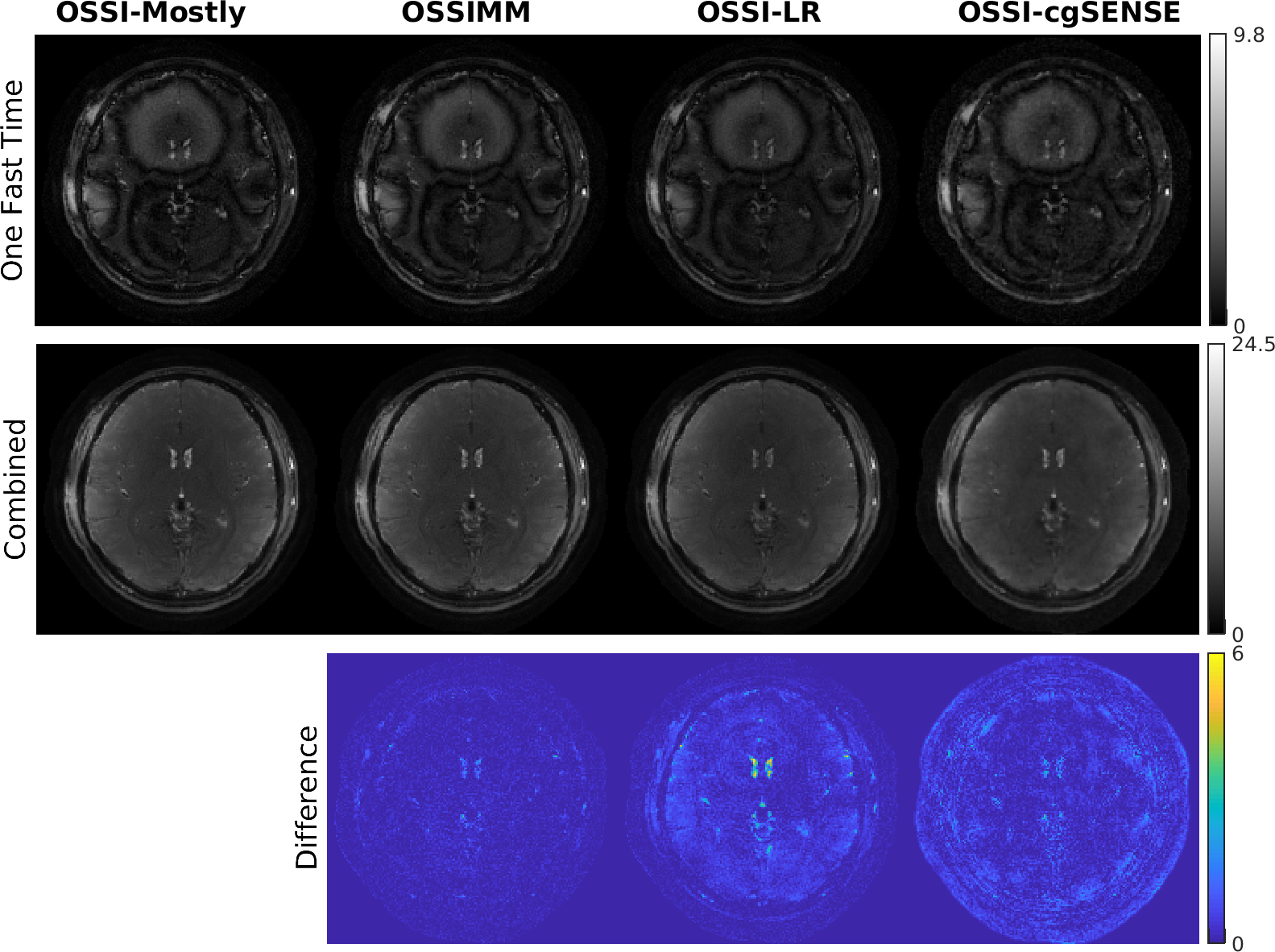}
  \end{minipage} \hfill
  \begin{minipage}[t]{0.27\textwidth}
  \vspace{0pt}
    \caption{
   Manifold, low-rank, and cgSENSE reconstructions for retrospectively undersampled OSSI data are compared to the mostly sampled reconstruction. The example fast time images present spatial variation in OSSI. OSSIMM outperforms other approaches with cleaner high-resolution details and less structure in the difference map.
    }\label{mf4}
  \end{minipage}
\end{figure*}

Figure \ref{mf4} compares retrospectively undersampled reconstructions to the mostly sampled reference.
OSSIMM reconstruction well preserves high-resolution structures in oscillatory fast time images and combined images,
and leads to less residual in the difference map than LR and cgSENSE approaches.

\begin{figure*}
  \begin{minipage}[t]{0.69\textwidth}
  \vspace{0pt}
     \includegraphics[width=\textwidth]{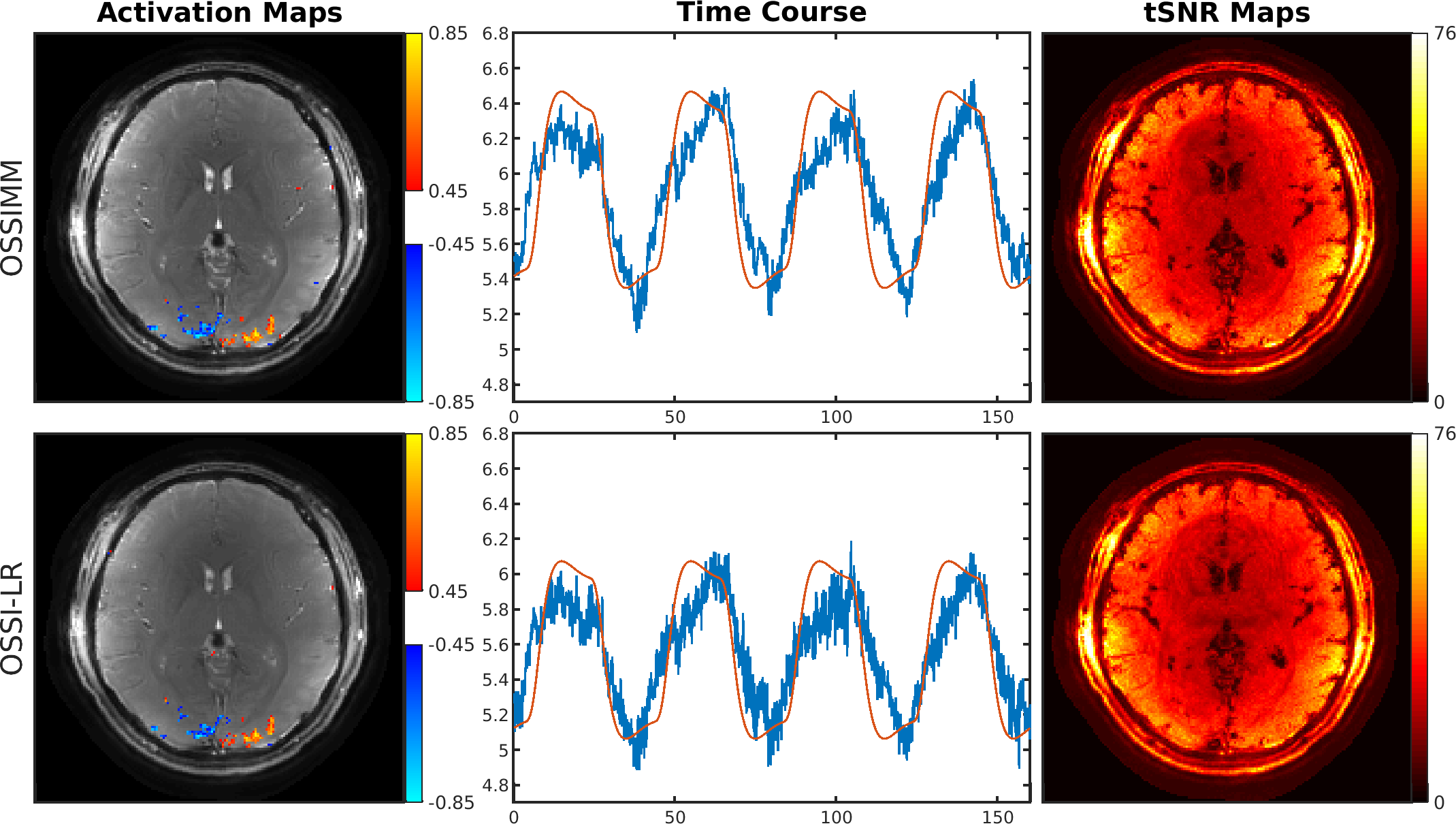}\\
     \includegraphics[width=\textwidth]{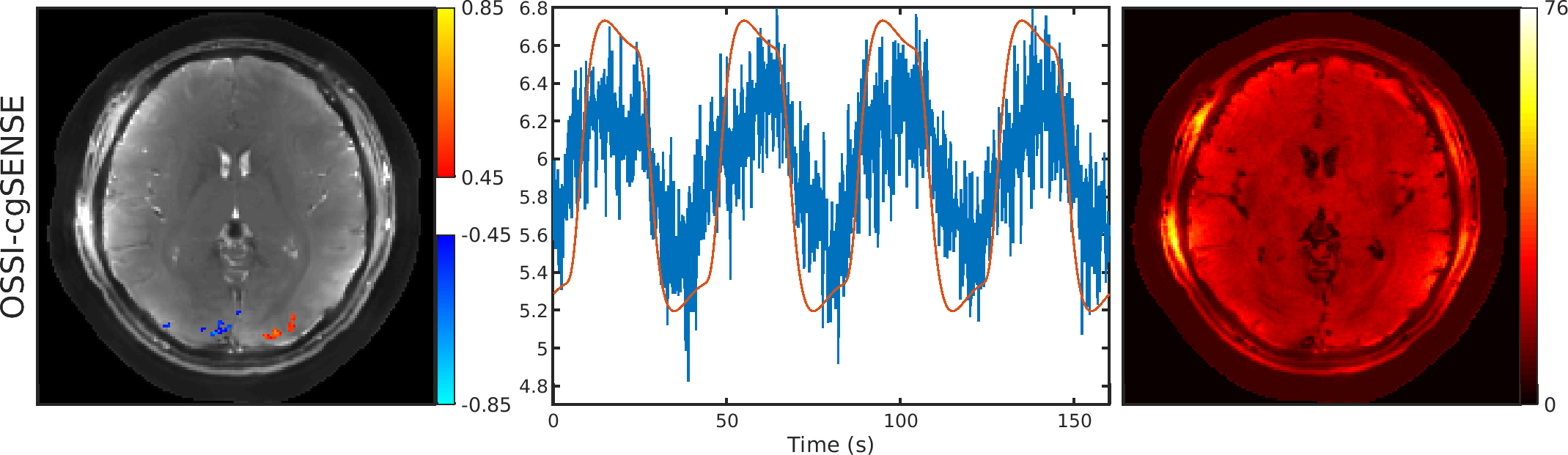}
\end{minipage} \hfill
\begin{minipage}[t]{0.27\textwidth}
  \vspace{0pt}
     \caption{
     Functional results for prospectively undersampled data with spatial resolution of 1.3 mm
     and temporal resolution of 150 ms.
     The proposed OSSIMM reconstruction provides an activation map
     with high-resolution background image and larger activated regions,
     and time course (reference waveform in red) and temporal SNR map with higher SNR than other methods.
     }
     \label{mf5}
   \end{minipage}
\end{figure*}

Figure \ref{mf5} presents prospectively undersampled reconstructions (temporal resolution = 150 ms)
using OSSIMM, LR, and cgSENSE.
OSSIMM demonstrates activation map with more activated voxels, time course with higher SNR,
and sharper tSNR map than other methods. 
The functional maps from the mostly sampled reconstruction (temporal resolution = 1.35 s)
are included in supplemental Fig.~\ref{msf3} for reference. 

\begin{figure*}
    \centering
    \begin{minipage}{0.495\textwidth}
        \includegraphics[width=\linewidth]{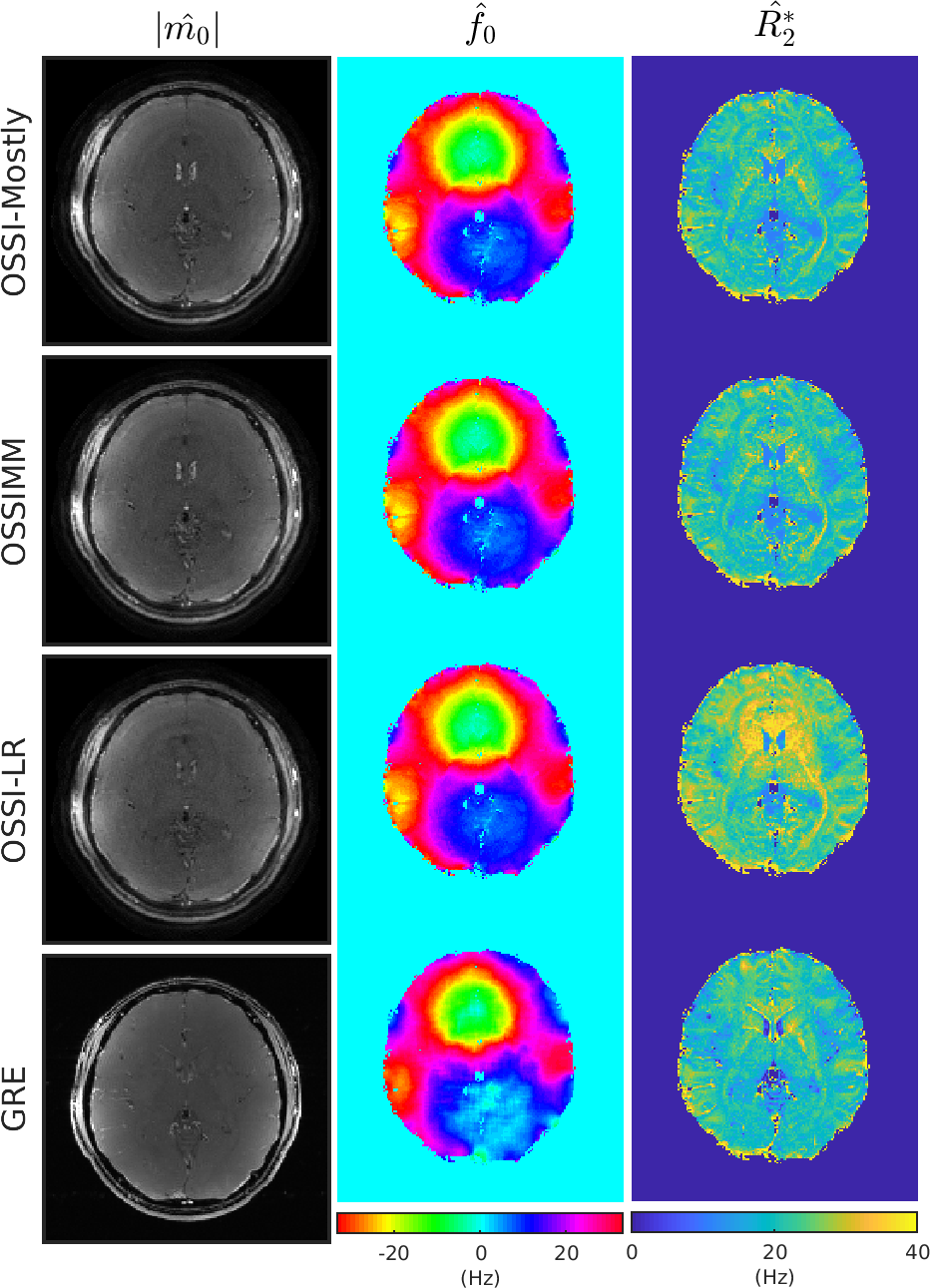}
        \includegraphics[width=\linewidth]{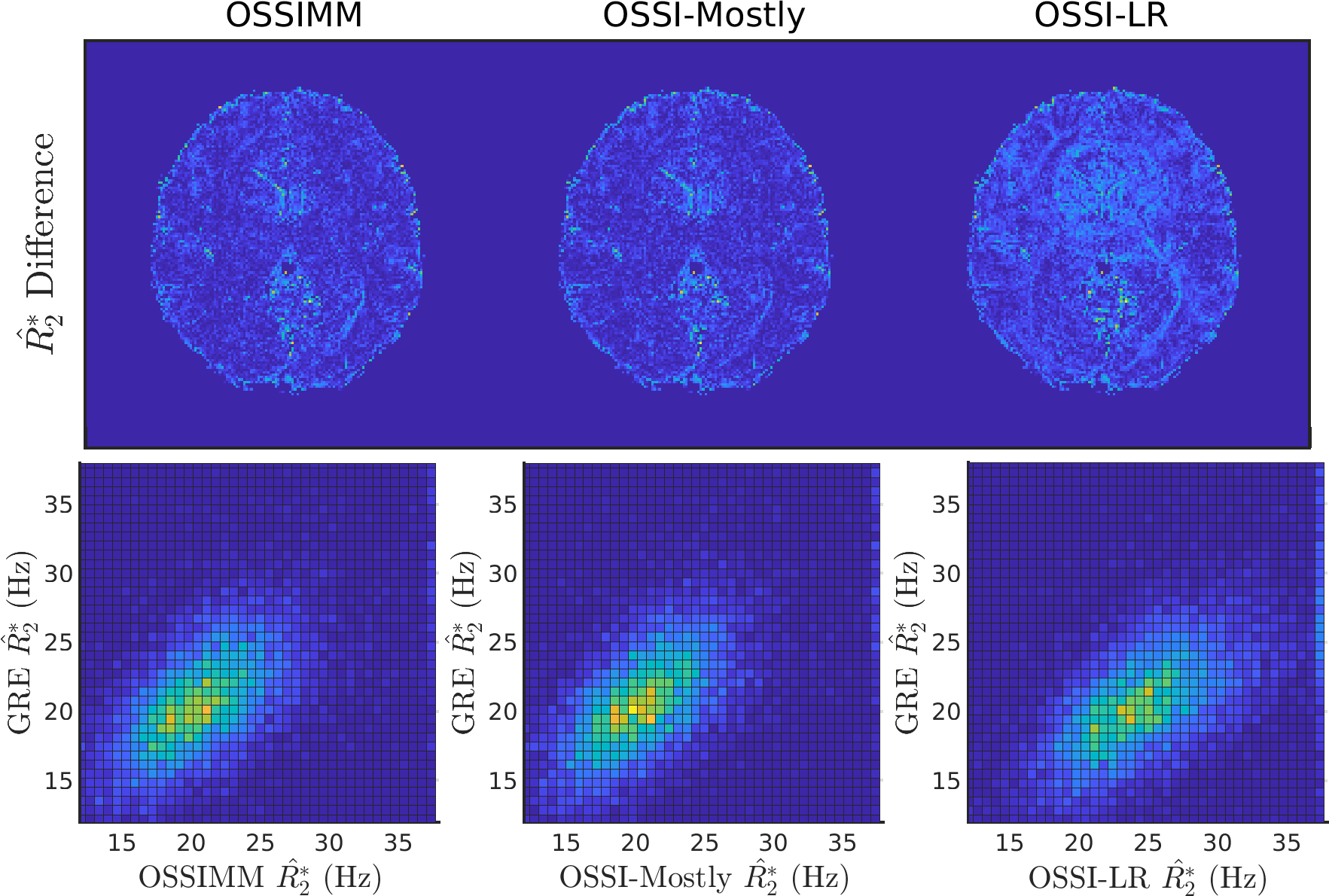}
        \caption{
   Retrospectively undersampled quantifications and comparison to multi-echo GRE estimates. OSSIMM presents similar results as the mostly sampled data. $\hat{R}_2^*$ difference maps (using GRE $\hat{R}_2^*$ as standard and of same color scale as $\hat{R}_2^*$ maps) and 2D histogram of $\hat{R}_2^*$ values show that OSSIMM provides comparable quantitative maps to GRE.
        }
        \label{mf6}
    \end{minipage}
    \hfill
    \begin{minipage}{0.495\textwidth}
        \includegraphics[width=\linewidth]{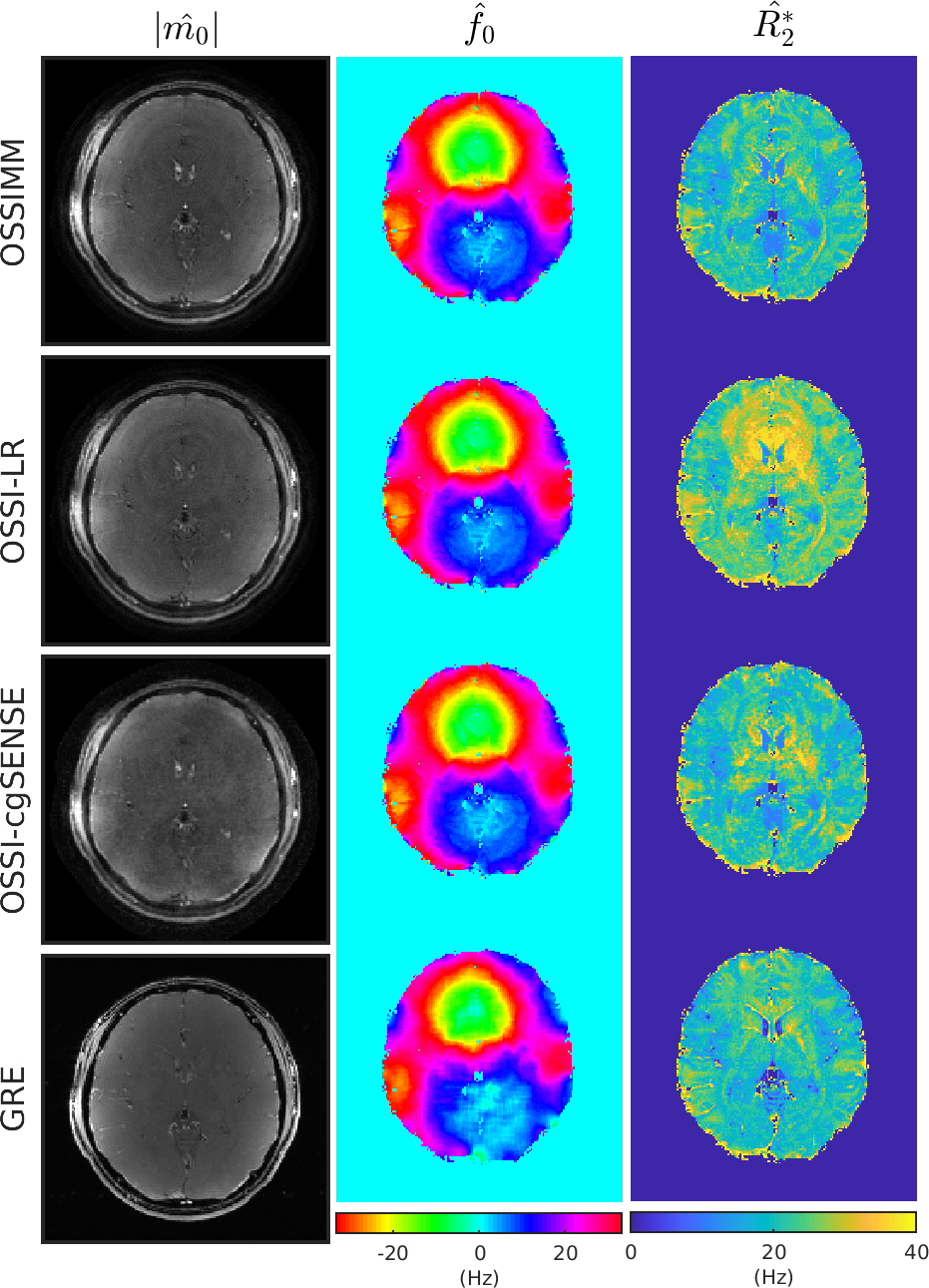}
        \includegraphics[width=\linewidth]{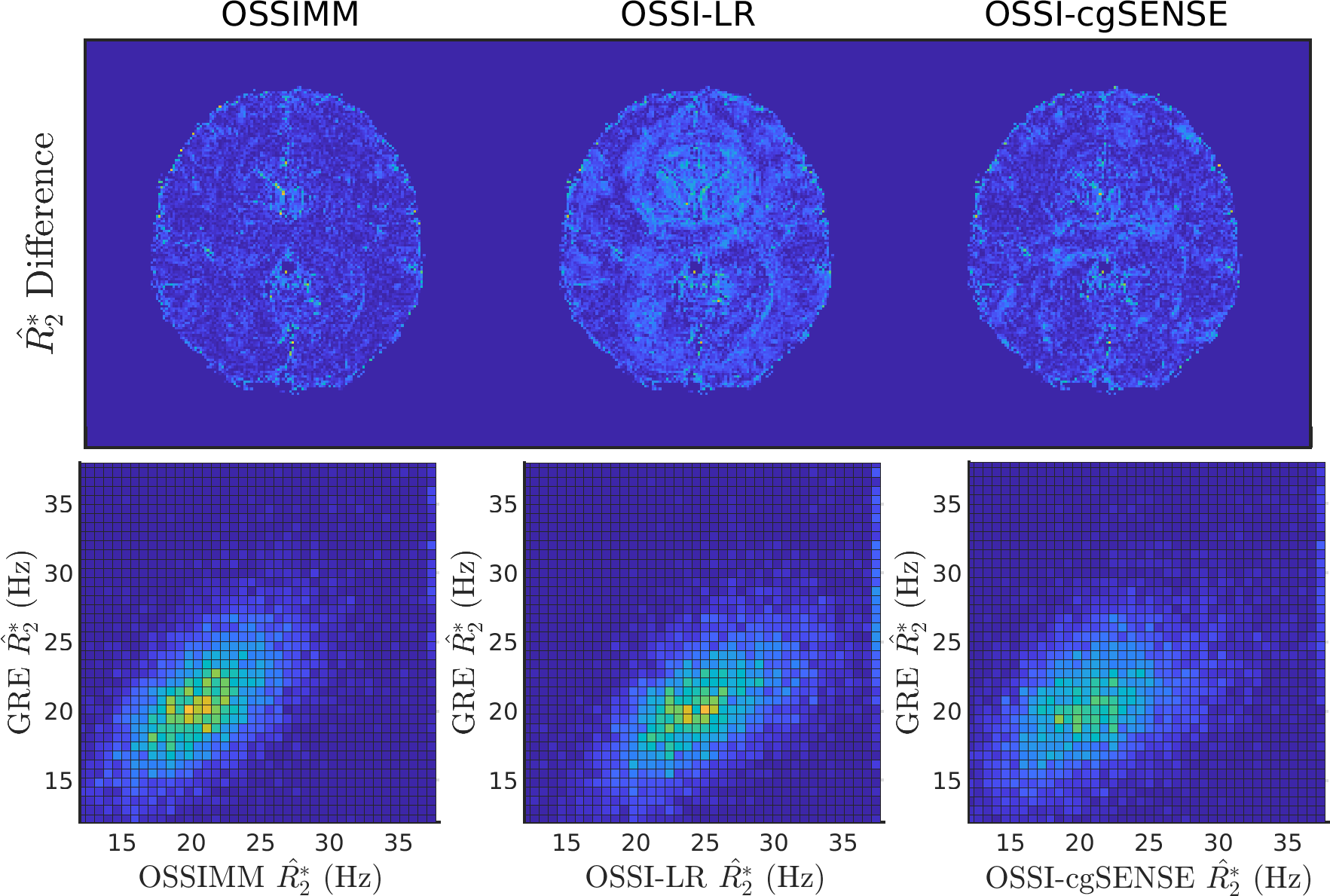}
        \caption{
   Prospectively undersampled quantifications compared to multi-echo GRE. OSSIMM results in reasonable parameter maps with 1.3 mm spatial resolution and a 150 ms acquisition time. OSSIMM also outperforms low-rank and cgSENSE reconstructions with less residual in the $\hat{R}_2^*$ difference map (same color scale as $\hat{R}_2^*$ maps).
        }
        \label{mf7}
    \end{minipage}
\end{figure*}

Figure \ref{mf6} gives retrospectively undersampled and mostly sampled OSSI quantification results
with comparison to multi-echo GRE.
OSSIMM with $12\times$ undersampling leads to $\hat{m}_0$, $\hat{f}_0$, and $\hat{R}_2^*$ estimates
that are almost identical to the mostly sampled case and have finer structures than OSSI-LR.
OSSIMM also provides comparable $\hat{R}_2^*$ maps to GRE
and demonstrates a similar distribution of $\hat{R}_2^*$ values within the brain as GRE
according to the 2D histogram.
Because of field drift and respiratory changes between different scans,
the OSSI-Mostly and OSSIMM $\hat{f}_0$ maps are close to GRE $\hat{f}_0$ but not exactly the same.

Figure \ref{mf7} compares prospectively undersampled quantification results to multi-echo GRE. 
OSSIMM enables high-resolution quantification of $m_0$, $R_2^*$ and $f_0$ with a 150~ms acquisition,
and yields parameter estimates more similar to GRE than LR and cgSENSE reconstructions.

\begin{figure*}
\centering
     \includegraphics[width=0.96\textwidth]{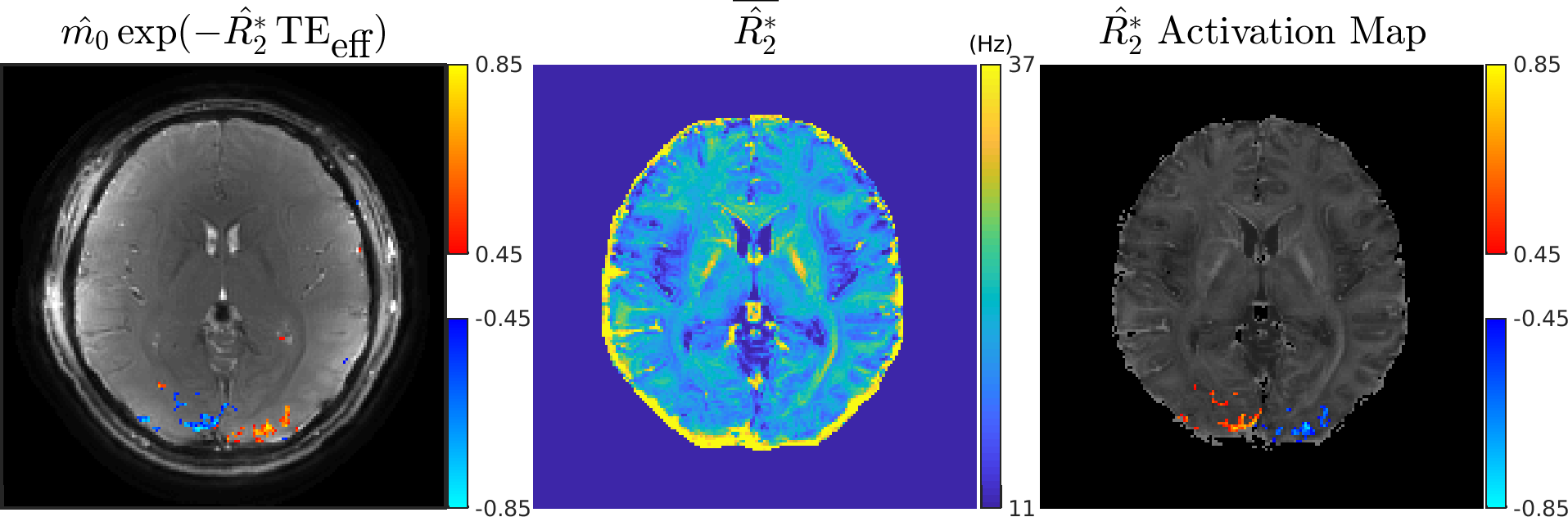}
   \caption{
   Activation maps from OSSIMM $\hat{m}_0$ and $\hat{R}_2^*$ with prospective undersampling demonstrating the dynamic quantification capacity of OSSIMM. Both time series of $\hat{m}_0\exp(-\hat{R}_2^*\,\textrm{TE}_\textrm{eff})$ (left) and $\hat{R}_2^*$ (right) almost fully recover the functional activation. The $\overline{\hat{R}_2^*}$ (middle) is the mean of $\hat{R}_2^*$ time series after skull stripping (without any other mask) and well preserves the $R_2^*$ contrast. 
   }\label{mf8}
\end{figure*}

The parameter maps in Figs. \ref{mf6} and \ref{mf7} are from a single set of $n_c = 10$ fast time images,
while OSSIMM jointly reconstructs undersampled measurements and quantifies physics parameters
for every 10 fast time images of the OSSI fMRI time course.
To demonstrate the dynamic quantification capacity of OSSIMM,
Fig.~\ref{mf8} shows activation maps
for $\hat{m}_0\exp(-\hat{R}_2^*\,\textrm{TE}_\textrm{eff})$
and $\hat{R}_2^*$,
where $\hat{m}_0$ and $\hat{R}_2^*$ are quantified using OSSIMM and prospectively undersampled data.
OSSI $\textrm{TE}_\textrm{eff}\,\approx$ 17.5 ms with a 2.6 ms actual TE \cite{ossi2019}. 

The activation maps based on $\hat{m}_0\exp(-\hat{R}_2^*\,\textrm{TE}_\textrm{eff})$ images
well preserves $R_2^*$ contrast of OSSI
and has the same activated regions as the activation map from 2-norm combined OSSI images (in Fig.~\ref{mf5}).
The activation map from $\hat{R}_2^*$ maps recovers the activation and reduces false positives
(negative activation in the positive activation region and vice versa).
The colors of the activation are the opposite of activation in Fig. \ref{mf5}
due to the negative correlation between $m_0\exp(-R_2^*\,\textrm{TE}_\textrm{eff})$ and $R_2^*$.
The mean $R_2^*$ map ($\overline{R_2^*}$) of the time series,
when compared to GRE, leads to a smaller RMSE value of 4.4 Hz.
The RMSE value = 3.7 Hz with a GRE $12 < \hat{R}_2^* < 38$ Hz mask.  

\begin{table}
\setlength{\tabcolsep}{4pt}
\renewcommand{\arraystretch}{1.6}
\caption{Human reconstruction and $R_2^*$ quantification evaluation for different sampling patterns and models}
\label{mt2}
\centering
\begin{tabular}{ccccc}
\hline\hline
& OSSIMM & OSSI-LR & OSSI-cgSENSE & OSSI-Mostly  
\\ \hline\hline
\multicolumn{5}{c}{Retrospectively Undersampled}
\\ \hline
$\hat{R}_2^*$ RMSE (Hz) 
& 5.1 & 6.6 & 5.4 & 5.1  
\\ \hline
Additional Mask 
& 4.5 & 6.1 & 4.9 & 4.5 
\\ \hline
\multicolumn{5}{c}{Prospectively Undersampled}
\\ \hline
$\hat{R}_2^*$ RMSE (Hz) 
& 4.9 & 6.7 & 5.5 & -     
\\ \hline
Additional Mask
& 4.3 & 6.4 & 5.0 & -     
\\ \hline
\# Activated Voxels
& 181 & 159 & 68 & -     
\\ \hline
Average tSNR
& 26.4 & 26.5 & 18.8 & -     
\\ \hline\hline
\end{tabular}
\end{table}

Table~\ref{mt2} summarizes quantitative evaluations of different sampling schemes and reconstruction models.
OSSI $\hat{R}_2^*$ RMSE values compared to GRE
for retrospectively (Fig. \ref{mf6})
and prospectively (Fig. \ref{mf7}) undersampling are presented.
As demonstrated by RMSE values with additional masking, 
OSSI RMSE decrease by about 0.5 Hz with the GRE $12 < \hat{R}_2^* < 38$ mask.
The last two rows of the table correspond to Fig.~\ref{mf5}
and are numbers of activated voxels and average tSNR within the brain for prospectively undersampled reconstructions.
The proposed OSSIMM jointly reconstructs high-resolution images
with more functional activation and parameter maps with smaller $\hat{R}_2^*$ RMSE than other approaches.

\section{Discussion}
We propose a novel manifold model OSSIMM that uses MR physics for the signal generation as the regularizer
for image reconstruction from undersampled k-space data.
The proposed model simultaneously provides high-resolution fMRI images
and quantitative maps of important MRI physics parameters. 

The proposed near-manifold regularizer has the advantage
of allowing for potential imperfections of the manifold model.
Instead of requiring the signal values to lie exactly on the manifold,
it provides a balance between fitting the fast-time images to the noisy k-space data and to the manifold. 
For reconstruction, OSSIMM outperforms low-rank and cgSENSE models
by providing more functional activation, without spatial or temporal smoothing. 

For quantification, OSSIMM dynamically tracks
$m_0$, $R_2^*$, and $f_0$ changes with a temporal resolution of 150 ms in our experiments.
The OSSIMM estimates  $\hat{m}_0\exp(-\hat{R}_2^*\,\textrm{TE}_\textrm{eff})$ or $\hat{R}_2^*$
contain most of the functional information of fMRI time series,
and may be well-suited for examining quantitative changes in longitudinal studies.
Moreover, OSSIMM quantification is faster than other quantification methods such as \cite{Wang2019}.
The manifold model and the near-manifold regularization
can be generalized to other sparsely undersampled datasets for joint reconstruction and quantification.  

There are multiple factors that contribute to slight mismatches between OSSI $\hat{R}_2^*$ and GRE $\hat{R}_2^*$.
We noticed that OSSI and GRE images were not exactly aligned
due to different gradient delays or the movement of the brain between different scans,
especially around the edge of the brain.
It is also possible that through-plane gradients change signals slightly differently between OSSI and GRE.
The OSSIMM implementation could be improved with a larger dictionary with a larger range of $R_2^*$ values
and finer spacing of the varying physics parameters.
The RF inhomogeneity in the brain may influence the accuracy of the dictionary fitting
due to inaccuracy of the flip angle.
Alternatively, one might include RF inhomogeneity in the dictionary and fit it as a nuisance parameter.

We have neglected the readout length effect for simplicity and have not performed field map correction for human data.
The field map correction improves quantification for resolution phantom,
but would increase computation for human fMRI time series.
One interesting extension would be to dynamically quantify $f_0$ and correct for field inhomogeneity
using the time-series of OSSI $\hat{f}_0$ maps.
Because OSSI $\hat{f}_0$ maps are in the range of [-33.3, 33.3] Hz,
we could use an initial estimate of $\hat{f}_0$ from two-echo GRE,
and dynamically update the initial $\hat{f}_0$ based on OSSI $\hat{f}_0$ changes along time as in \cite{Olafsson2008}.

We believe that the reconstruction performance can be further improved
with spatial-temporal modeling of OSSI fMRI image series.
We will combine OSSIMM with the patch-tensor low-rank model \cite{tensor}
to exploit different aspects of prior information (linear and nonlinear),
enlarge the capacity of regularization,
and enable more aggressive undersampling.
We will also extend the OSSIMM dynamic quantification to 3D fMRI.
Because a known $\hat{T}_2$ map can be helpful for $\hat{R}_2^*$ estimation,
one might considering modifying the OSSI sequence as in \cite{Wang2019}
with slowly varying flip angles and other changes to simultaneously quantify $T_2$ and $T_2’$.

\section{Conclusion}

This paper proposes OSSIMM,
a novel reconstruction and quantification model for nonlinear MR signals.
With a factor of 12 undersampling and without spatial or temporal smoothing,
OSSIMM outperforms other reconstruction models
with high-resolution structures and more functional activation.
OSSIMM also provides dynamic $R_2^*$ maps that are comparable to GRE $\hat{R}_2^*$ maps
with a 150 ms temporal resolution. 

\section*{Acknowledgment}

The authors would like to thank Dr. Amos Cao for important discussions on signal modeling and Dinank Gupta for helping with spin-echo imaging. 


\newpage
\onecolumn
\renewcommand{\thefigure}{S\arabic{figure}}

\section*{\LARGE Supplemental}
\vspace*{1cm}

\begin{figure}[!htbp]
\centering
     \includegraphics[width=0.92\textwidth]{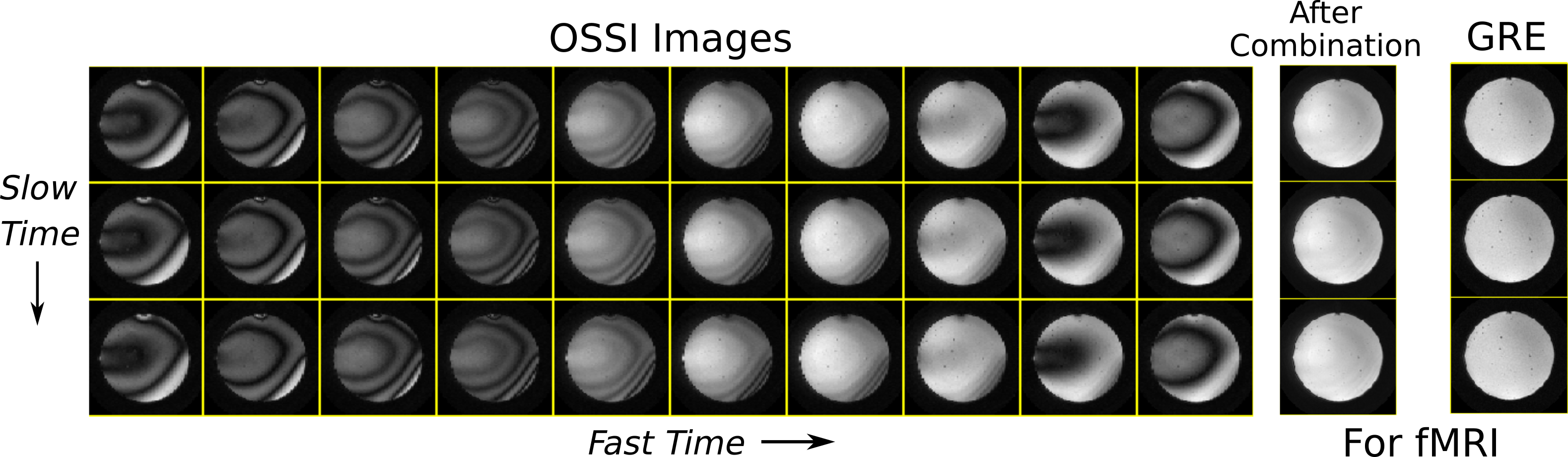}
     \caption{
OSSI images with periodic and nonlinear oscillation patterns are structured along ``fast time” and ``slow time”. Every $n_c$ fast time images can be 2-norm combined to generate fMRI images that have comparable $T_2^*$-sensitivity as standard GRE fMRI.
     }
     \label{msf0}
\end{figure}

\begin{figure}[!htbp]
\centering
     \includegraphics[width=0.45\textwidth]{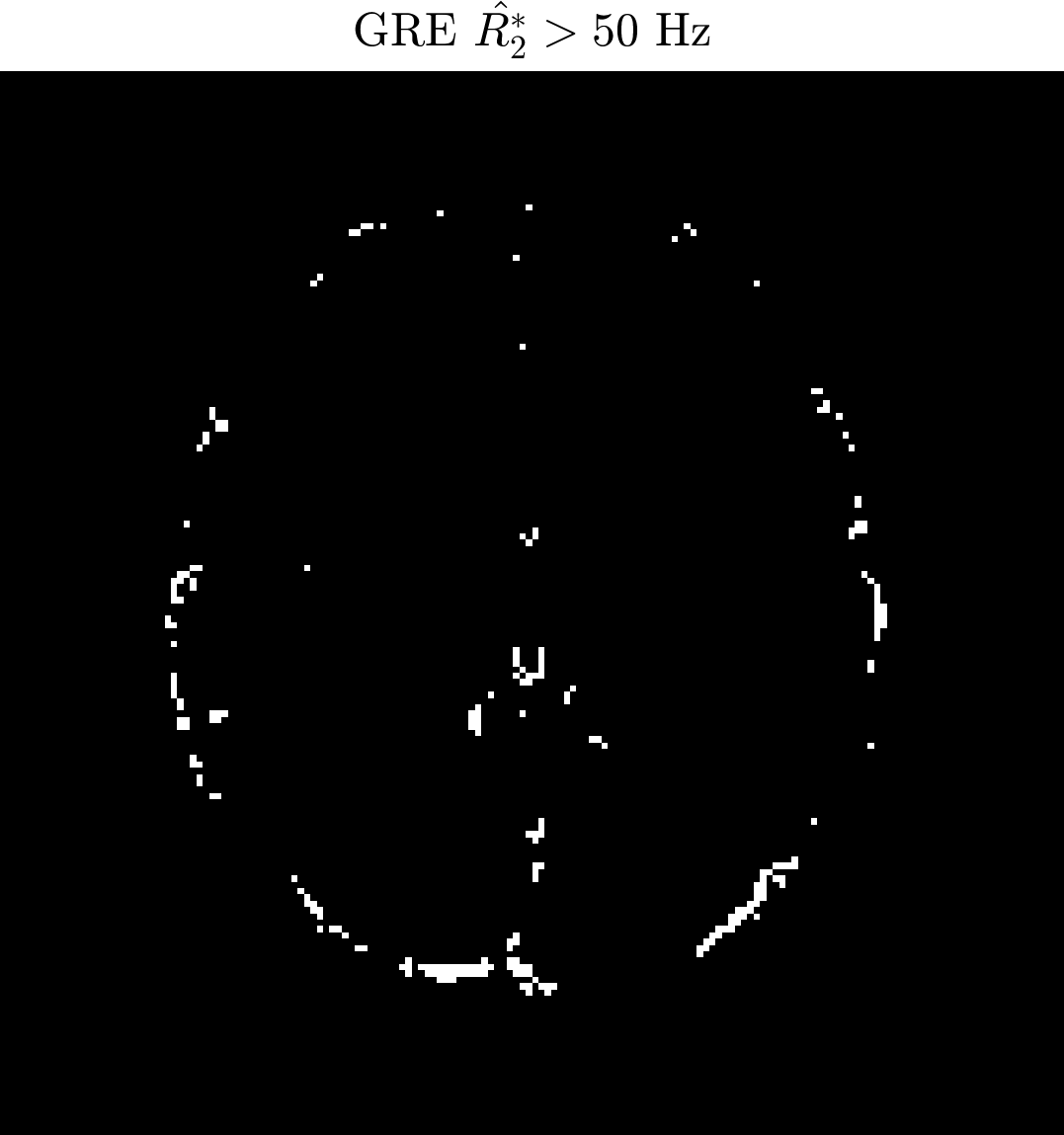}
     \caption{
Most voxel locations with GRE $\hat{R}_2^*\, >$ 50 Hz are around the edges of the brain.
     }
     \label{msf1}
\end{figure}

\begin{figure}[!htbp]
\centering
    \includegraphics[width=0.95\textwidth]{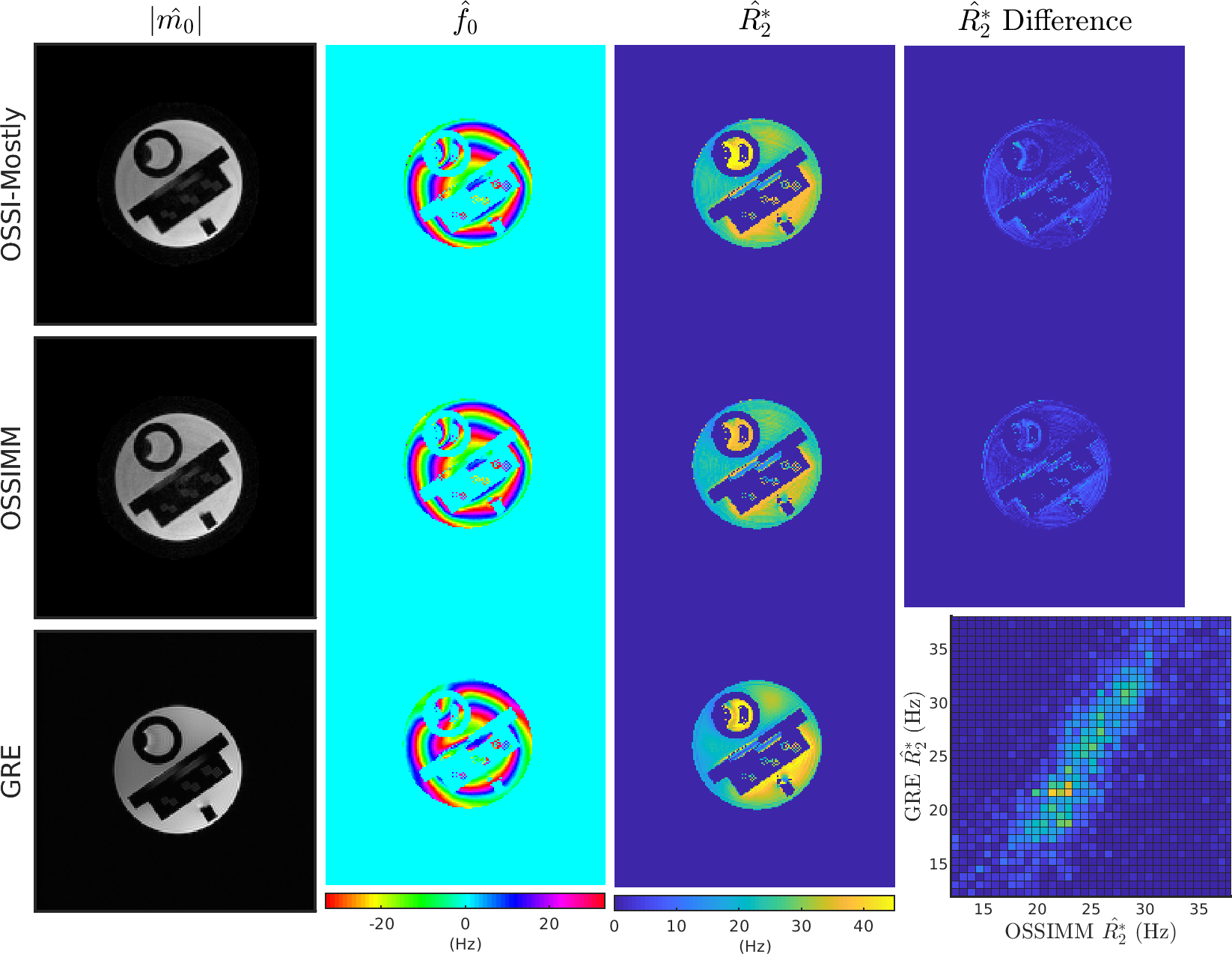}
   \caption{
   Phantom quantification of $m_0$, $f_0$, and $R_2^*$ from mostly sampled OSSI data,
   retrospectively undersampled OSSI data (reconstructed and quantified using OSSIMM with a known $\hat{T}_2$ map),
   and multi-echo GRE. The $\hat{m}_0$ estimates are on arbitrary scales.
   The GRE $\hat{R}_2^*$ map is used as the standard for difference map calculation.
   The $\hat{R}_2^*$ maps and $\hat{R}_2^*$ difference maps use the same color scale.
   The 2D histogram (bottom right) compares OSSIMM and GRE $\hat{R}_2^*$ within the 12-38 Hz range.
   OSSI $\hat{R}_2^*$ and GRE $\hat{R}_2^*$ have similar contrasts.
   }\label{msf2}
\end{figure}

\begin{figure}[!htbp]
\centering
     \includegraphics[width=0.95\textwidth]{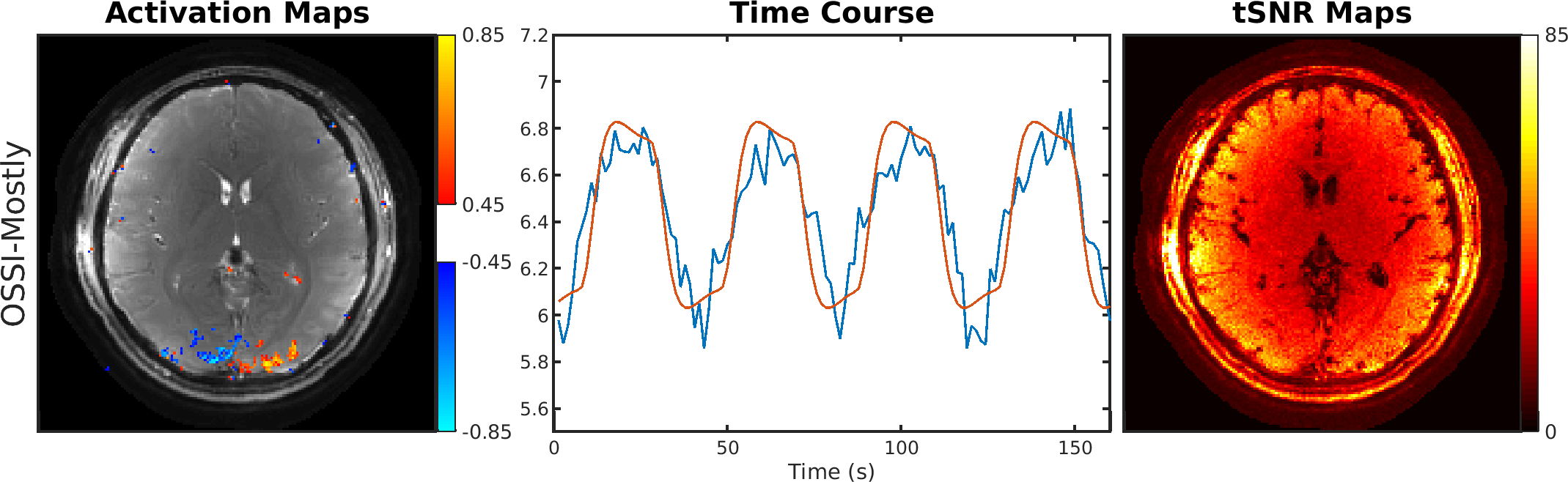}
     \caption{
     Functional results for mostly sampled data with spatial resolution of 1.3 mm and temporal resolution of 1.35 s. The number of activted voxels is 236, and the average temporal SNR within the brain is 31.3.
     }
     \label{msf3}
\end{figure}

Figure~\ref{msf0} illustrates OSSI ``fast time" and ``slow time".
Figure~\ref{msf1} demonstrates voxel locations with GRE $\hat{R}_2^*\, >$ 50 Hz.
Figure~\ref{msf2} presents phantom quantification results,
and OSSI quantitative maps that were calculated with a known $\hat{T}_2$ map.
Figure~\ref{msf3} presents fMRI results for mostly sampled human data.

\end{document}